\documentclass[11pt]{article}
\usepackage{latexsym, amsmath, xcolor, graphicx, amssymb, color,booktabs,mathrsfs}
\usepackage[final]{pdfpages}
\usepackage[left=1in,right=1in,top=1in,bottom=1in]{geometry}
\usepackage{biblatex}
\usepackage{float}
\usepackage{caption}
\usepackage{subcaption}
\begin{document}

%%%% Article title to be placed here
\title{Postural Orthostatic Tachycardia Syndrome explained using a baroreflex response model}

\author{%%%% Author details
Justen Geddes$^{1}$, Johnny T Ottesen$^{2}$,\\ Jesper Mehlsen$^{3}$ and Mette S. Olufsen$^{1}$}

\maketitle

% %%%%%%%%% Insert author address here

% %%%% Subject entries to be placed here %%%%
% \subject{Computational Physiology}

% %%%% Keyword entries to be placed here %%%%
% \keywords{POTS, baroreflex, mathematical modeling}

% %%%% Insert corresponding author and its email address}
% \corres{Mette S Olufsen\\
% \email{msolufse@ncsu.edu}}

\begin{abstract}
Recent studies have shown that Postural Orthostatic Tachycardia Syndrome (POTS) patients have abnormal low frequency $\approx 0.1$ Hz blood pressure and heart rate dynamics. These dynamics are attributed to the baroreflex and can give insight into the mechanistic causes which are the basis for proposed subgroups of POTS. In this study we develop a baroreflex model replicating the low-frequency dynamics observed in POTS patient data as well as represent subgroups of POTS in our model. We utilize signal processing to quantify the effects that model parameters have on low-frequency oscillations. Results show that key physiological parameters that represent the hypothesized causes are central to our model's ability to reproduce observed dynamics from patient data.
\end{abstract}

\noindent $^{1}$ Department of Mathematics North Carolina State University, Raleigh, NC, 27695, USA\\
$^{2}$ Department of Science and Environment, Roskilde University, Denmark\\
$^{3}$ Section for Surgical Pathophysiology, Rigshospitalet, Denmark
\section{Introduction}

Postural Orthostatic Tachycardia Syndrome (POTS) is characterized by the presence of tachycardia upon the transition to an upright position in addition to a history of persistent (at least six months) symptoms, absence of orthostatic hypotension, and absence of other condition provoking sinus tachycardia \cite{fedorowski2019postural, freeman2011consensus, mar2020postural}.  Symptoms include mild to severe brain fog, palpitations, visual blurring, and/or dizziness. Since POTS is a phenotype and not a specific disease, it is difficult to identify the compromised mechanisms. This is partly due to POTS' numerous potential causes, including dehydration, neuropathy, or the presence of agonistic antibodies binding to specific adrenergic receptors \cite{li2014autoimmune, mar2020postural}. POTS is typically diagnosed by examining heart rate and blood pressure in response to a postural challenge, such as head-up tilt (HUT) or active standing \cite{raj2006postural}. These signals are measured continuously and re-ported along with a description of symptoms, yet diagnosis primarily relies on a single quantity - tachycardia (heart rate increase $> 30$ bpm, $> 40$ bpm for adolescents) \cite{raj2006postural,singer2012postural}.

Reliance on postural tachycardia to diagnose POTS is problematic as the single measure does not provide insight into the mechanistic causes underlying the syndrome.  Recent studies \cite{Geddes2020Osc,stewart2015oscillatory,medow2014altered}, have observed that not only do POTS patients exhibit tachycardia, increasing heart rate higher than normal in response to postural change, they also experience increased 0.1 Hz heart rate and blood pressure oscillations. Curently, clinical diagnosis only includes one POTS group, but as suggested by Mar, Raj, and Fedorowski \cite{mar2020postural, fedorowski2019postural}, POTS may have three phenotypes: (1) neuropathic POTS caused by neuropathy in the vascular beds, particularly in the lower body; (2) hypovolemic POTS attributed to low fluid volume in the body and (3) hyperadrenergic POTS characterized by high levels of circulating norepinephrine during postural change inducing an exaggerated sympathetic response (Grubb08).  Diagnosis of these subtypes typically involves multiple tests as phenotypes can be challenging to identify from heart rate and blood pressure response to a single HUT test. 

It is known that POTS patients typically experience compromised baroreflex function \cite{mar2020postural, Geddes2020Osc}. Several hypotheses have been put forward suggesting what parts of the system are compromised, though it is difficult to determine how each factor impacts dynamics. As a result, most patients receive a series of tests to examine their dynamic response. This study uses a mathematical model to investigate how the system responds when parameters associated with each phenotype are varied. More insight into how the system reacts to a specific change may reduce the number of tests needed for accurate diagnosis. 

For healthy people, the baroreflex system operates via negative feedback modulating sympathetic and parasympathetic nerve activity mitigating blood pressure changes. Stretch receptors in the aortic arch and carotid sinus detect changes in blood pressure modulating firing rate in the afferent vagal nerve, which sends signals to the nucleus tractus solitarius (NTS). From here, the signals are transmitted via the efferent sympathetic and parasympathetic nerves. Heart rate is modulated by changes in the firing of both sympathetic and parasympathetic nerves, while the sympathetic nervous system primarily modulates the peripheral vascular resistance and cardiac contractility. At rest, sympathetic activity is low ($\approx 20\%$ of its maximum), while the parasympathetic activity is high ($\approx 80\%$ of its maximum) \cite{korner1976reflex}. In response to a decrease in blood pressure, the afferent signaling is inhibited, leading to parasympathetic withdrawal and sympathetic stimulation increasing in heart rate, cardiac contractility, and peripheral resistance \cite{Boron}. Numerous studies have examined baroreflex signaling \cite{cevese2001baroreflex, deboer1987hemodynamic, medow2014altered}, and it has been established that blood pressure and heart rate are controlled by negative feedback with a resonance frequency of approximately 0.1 Hz. This response is easily distinguished from heart rate with a frequency of 1 Hz and respiration, which oscillates with a frequency of $0.2-0.3$ Hz \cite{deboer1987hemodynamic}. 

As noted earlier, several recent studies have examined the magnitude, and phase of the low frequency $\approx 0.1$ Hz) blood pressure and heart rate oscillations in POTS patients \cite{Geddes2020Osc, medow2014altered, stewart2015oscillatory}. The studies by Stewart et al. \cite{stewart2015oscillatory} and Medow et al. \cite{medow2014altered} used Transcranial Doppler measurements of cerebral blood flow and finger arterial plethysmography to analyze blood flow and heart rate oscillations in response to a postural challenge. Using auto-spectral and transfer function analysis, they reported that increased low-frequency oscillations in arterial pressure led to increased oscillations in cerebral blood flow, which they suggest may be responsible for the ``brain fog" experienced by many POTS patients. These results agree with our findings using empirical mode decomposition to examine blood pressure and heart rate signals measured during HUT from females diagnosed with POTS.  We found that the magnitude of the ~ 0.1 Hz heart rate (HR) and blood pressure (BP) oscillation was increased during HUT and that the instantaneous phase difference between low-frequency HR and BP signals is shorter in POTS patients than in control subjects, at rest and during HUT \cite{Geddes2020Osc}. These studies indicate that POTS patients have compromised baroreflex and that it is likely that both the sympathetic and parasympathetic branches are compromised. Several studies \cite{bryarly2019postural, fedorowski2019postural, mar2020postural} discuss what parts of the system may be compromised, but it is difficult to obtain direct measures explaining how specific pathophysiology impact heart rate and blood pressure dynamics.

One way to gain more insight into how a specific pathophysiology impacts the system dynamics is by building mechanistic models and comparing signals from controls and POTS patients. Numerous studies have examined the baroreflex feedback dating back to studies by Bronk and Stella \cite{bronk1934response} and by Landgren et al. \cite{landgren52} who built a mechanical apparatus to study how changes in pressure modulate the firing of the baroreceptor nerves in the carotid artery in rabbits and cats. Data were analyzed using a simple mathematical model. This study was followed by a series of studies \cite{spickler67, Beneken1967physical, guyton72, srinivasan72} using modeling to relate blood pressure and heart rate. The study by Mahdi et al. \cite{mahdi13} gives an overview of several early models. The most notable results are by Beneken and DeWitt, who modeled the baroreflex as a transfer relation with two ``regions" - the first associating large changes in pressure with a short time-constant, and the second smaller changes in pressure with a larger time-constant, and the comprehensive model by Guyton \cite{guyton72} explaining blood pressure control. Since then, numerous researchers have examined various aspects of the baroreflex system, including several contributions by Ursino et al. (e.g., \cite{ursino98, ursino03}) describing the principal baroreflex mechanisms. These early studies focused on describing mechanisms underlying the baroreflex function, while the more recent studies focus on calibrating models to data. An example is the study by Bugenhagen et al. \cite{bugenhagen2010identifying} that used blood pressure as an input to fit spontaneous baroreflex regulation of heart rate in salt-sensitive Dahl rats.

These models must be translated to examine the response to typical postural challenges such as HUT and active standing imposed to understand human pathophysiology. Several studies have examined the response to orthostatic stress challenges, e.g., \cite{heldt02, olufsen05, ellwein08, batzel07, ottesen11, williams2014patient, matzuka2015using}. For example, the model by Olufsen et al. \cite{olufsen06} used heart rate as an input to predict the blood pressure response to active standing. The model demonstrates that model predictions can match a healthy young adult's blood pressure by estimating patient-specific cardiovascular parameters modulating peripheral vascular resistance and vascular compliance.  In \cite{williams2014patient} this approach was adapted to study the response to HUT. In Matzuka et al. \cite{matzuka2015using} parameter estimation was carried using Kalman Filtering, while the study by Williams et al. \cite{williams2019optimal} and Matzuka et al. \cite{matzuka2015using} used optimal control theory to estimate model parameters.

While these studies all captured variations in response to a postural change, none tested the frequency of baroreflex changes examining the power of the characteristic 0.1 Hz oscillations. To our knowledge, only a few studies have attempted to test if dynamical systems models dis-play 0.1 Hz oscillations. The study by Heldt et al. \cite{heldt2000computational} built a model predicting low-frequency oscillations in astronauts undergoing a sit-to-stand test using a baroreflex control model. They found that the low-frequency oscillations emerging from their model did not persist after the transition from sit-to-stand. Another attempt was made by Hammer and Saul \cite{hammer2005resonance}, who used an open loop baroreflex model to predict the response to a postural change. This model uses arterial blood pressure as an input to predict heart rate. While this model examines the 0.1 Hz oscillations, it does not study how the response changes in time; instead, it quantifies stability at fixed operating points responsible for low-frequency oscillations. More recently, Ishbulatov et al. \cite{ishbulatov2020mathematical} use a closed-loop baroreflex model to replicate low-frequency aspects of patient data during a passive HUT test. This study analyzes how a healthy human body adapts to an orthostatic challenge. However, this model is complex and does not study the response in POTS patients. 

To remedy the shortcomings of these previous studies, we use a simple closed-loop differential equations model without delays to examine temporal and frequency baroreflex response to HUT for POTS patients. 

To our knowledge, no previous studies have combined a mechanistic model with signal analysis to explain the emergence and modulation of the low-frequency oscillations for POTS patients. To do so, we develop a systems-level baroreflex model to explain the low-frequency dynamics observed in POTS patients both at rest and during HUT. We use simulations to display how the three POTS phenotypes identified by \cite{mar2020postural} can be encoded in the model and how these possibly affect blood pressure and heart rate dynamics. Our model is formulated using a simple closed-loop 0D cardiovascular model, with basic first-order control equations representing the baroreflex regulation. We analyze our model using signal processing techniques and study the effects of critical model parameters that correspond to the physiological abnormalities that cause each POTS phenotype. Results indicate that changes in clinically relevant parameters can result in the emergence of low-frequency oscillations with amplitude equal to that observed in POTS patient data from our previous study \cite{Geddes2020Osc}. Discussion of our results focuses on clinical implications and motivation of future studies.

\section{Methods}

This study develops a closed-loop 0D model describing the emergence of low-frequency ($\approx$ 0.1 Hz) oscillations observed in POTS patients. The model is parameterized to match average blood pressure and heart rate signals measured during HUT. Simulation results are depicted with characteristic POTS data. The model is simulated both at rest and during HUT with varying parameters that differentiate the three phenotypes suggested by Mar and Raj \cite{mar2020postural}.

\begin{figure}[ht!]
    \centering
    \includegraphics[width = \textwidth]{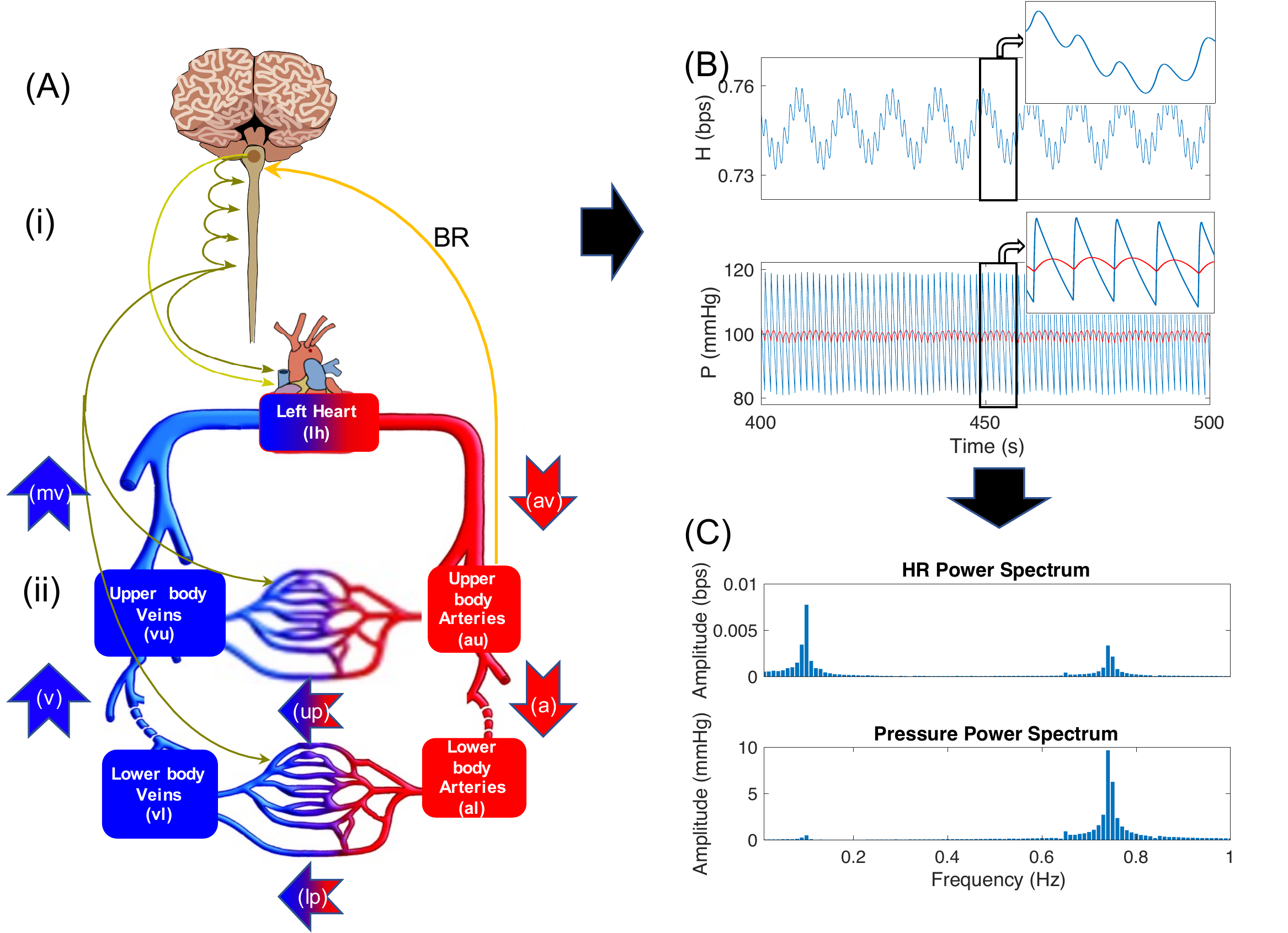}
    \caption{(Ai) Hemodynamics is controlled by the baroreflex system, which senses changes in the upper body arteries (lumping thoracic and carotid baroreceptors). Afferent signals from baroreceptor neurons are integrated into the brain and transmitted via sympathetic and parasympathetic neurons regulating heart rate, peripheral vascular resistance, and left heart elastance. (Aii) The systemic circulation is represented by compartments lumping upper ($au$) and lower ($al$) body arteries, upper ($vu$) and lower ($vl$) body veins, and the left heart ($lh$). Flow ($Q$) through the aortic valve ($av$) is transported from the left heart to the upper body arteries. From here, it is transported in the arteries ($a$) to the lower body arteries and through the upper body peripheral vasculature ($up$) to the upper body veins. A parallel connection transports flow through the lower body peripheral vasculature ($lp$). From the lower body venous flow ($v$) is transported to the upper body veins and finally via the mitral valve ($mv$) back to the left heart. Each compartment representing the heart or a collection of arteries or veins has pressure ($P$), volume ($V$), and elastance ($E$). Pumping of the heart is achieved by assuming that left heart elastance ($E_{lh}(t)$) is time-varying. (B) Model predictions of heart rate ($H$ (bps), top panel) and upper body arterial pressure ($P_{au}$ (mmHg), lower panel). The blue line shows pulsatile blood pressure and the red the mean pressure ($P_m$). 5-second sections of each signal are shown in the overlaid subpanels. (C) Frequency spectra of time-series data ($H$ top, $P_{au}$ bottom) shown in (B).}
    \label{fig:System}
\end{figure}

Similar to previous studies \cite{williams2014patient,ottesen2013development}, we predict blood flow and pressure in the systemic circulation using an electrical circuit model with five compartments, including the upper and lower body arteries and veins, and the left heart (see Figure \ref{fig:System}Aii). The baroreflex is incorporated via negative feedback control-equations predicting the effector response (heart rate, vascular resistance, and cardiac contractility) as functions of mean arterial pressure (Figure \ref{fig:System}Ai). The magnitude and phase of the low-frequency oscillations generated by the baroreflex are extracted using discrete Fourier transform, analyzing computed heart rate and blood pressure signals. 

Computations are first conducted in the supine position, followed by HUT simulated by shifting blood from the upper to the lower body. We demonstrate the importance of incorporating heart rate variability by adding uniformly distributed noise to predictions of heart rate, and discuss how phenotypes suggested by Mar and Raj can be simulated.

\subsection{Data}

The model simulations are qualitative in nature and meant to illustrate how changing system properties impact dynamics. But test if the outcome of our simulations make sense in terms of physiological behavior, we included blood pressure and heart rate measurements (extracted from \cite{Geddes2020Osc}) from two representative subjects: a control and a POTS patient.

Measurements from these people include  continuous ECG and upper arterial blood pressure measurements extracted at rest for 5 minutes and then for 5 additional minutes after the HUT onset. Heart rate is extracted from intervals between consecutive RR waves obtained from a 3-lead ECG, and continuous blood pressure measurements are obtained using a Finapress device (Finapres Medical Systems BV, Amsterdam, Netherlands). Blood pressure and ECG signals are sampled at 1000 Hz. Heart rate is extracted from the high-resolution ECG measurements as the inverse distance between consecutive RR intervals. Blood pressure and heart rate signals are sub-sampled to 250 Hz, after which the Uniform Phase Empirical Mode Decomposition \cite{Geddes2020Osc} is used to extract the magnitude and phase of 0.1 Hz oscillations.

The afferent input to the baroreflex control model assumes that blood pressure is measured at the level at the carotid barorecptors (above the center of gravity). Therefore blood pressure data, measured at the level of the heart, is adjusted by subtracting the effect of gravity as described in our previous study \cite{williams2014patient}.

\subsection{Cardiovascular model}

We employ an electrical circuit analogy to predict blood flow (analogous to current), pressure (analogous to voltage), and volume (analogous to charge) in the systemic circulation represented by five compartments, including the upper ($u$) and lower ($l$) body arteries ($a$) and veins ($v$), and the left heart ($lh$). Each compartment is  quantified by its volume ($V(t)$ ml) and pressure ($P(t)$ mmHg), while flow  $Q(t)$ (ml/s) exists between compartments. Figure \ref{fig:System} depicts the model and Table \ref{tab:State table} lists the dependent cardiovascular variables. 

To ensure flow conservation, for each compartment ($i = lh, au, al, vl, vu$), the change in volume is computed as the difference between flow into and out of the compartment, 
\begin{equation}
    \frac{dV_i}{dt} = Q_{in} - Q_{out},
\end{equation}
where $Q_{in}$ denotes the flow into, and $Q_{out}$ denotes the flow out, of compartment $i$. Ohm’s law relates flow to pressure and the resistance ($R$, mmHg s/ml) between compartments $(i-1)$ and ($i$), 
\begin{equation}
    Q_i = \frac{P_{i-1}-P_i}{R_i}.
\end{equation}
For each arterial compartment and upper venous compartment $i$, pressure and volume are related using the linear relation 
\begin{equation}\label{eqn:compliance pressure eq vol}
    P_i-P_{ui} = E_i(V_i-V_{ui}),
\end{equation}
where $V_{ui}$ is the unstressed volume, $E_i$  is the elastance (reciprocal of compliance, analogous to capacitance), and $P_{ui}=0$ is the unstressed pressure.

Given that pressure changes significantly on the venous side, in particular in the lower venous compartment during HUT, as suggested by Hardy et al. \cite{hardy1982pressure} we employ a nonlinear relation between lower venous pressure and volume given by 
\begin{equation} \label{eqn:Hardy pressure}
     P_{vl} = \frac{1}{m_{vl}} \log \Big( \frac{V_{Mvl}}{V_{Mvl} - V_{vl}}\Big)
 \end{equation}
 where $m_{vl}$ is a parameter that relates nominal pressure, volume ($V_{vl}$) and maximal volume ($V_{Mvl}$) \cite{pstras2017mathematical}.
 
The pumping of the heart is achieved by introducing a time-varying elastance function of the form
\begin{equation}
    E_{lh}(t) = \begin{cases} \frac{E_S-E_D}{2} \Big(1-\cos(\frac{\pi t}{T_S}) \Big) + E_D & 0 \leq t \leq T_S\\
    \frac{E_S-E_D}{2} \Big(\cos\big(\frac{\pi(t-T_S)}{T_D}\big)+1\Big) +E_D & T_S\leq t \leq T_S +T_D \\
    E_D & T_S + T_D \leq t \leq T,
    \end{cases}
\end{equation}
where $E_S, E_D, T_S,$ and $T_D$ denote the end systolic and end diastolic elastance, the time for systole and diastole, respectively.

The timing parameters $T_S$ and $T_D$ are determined as functions of the length of the previous cardiac cycle (the RR interval). By combining the prediction of the length of the QT interval from \cite{akhras1981relationship,kovacs1985duration} and the ratio of cardiac mechanical contraction to relaxation from \cite{janssen2010kinetics}, we get
\begin{equation}
    T_S = 0.45\Big(c_1+\frac{c_2}{RR}\Big), \ \ \ \ \ T_D = 0.55\Big(c_1+\frac{c_2}{RR}\Big),
\end{equation}
where $c_1 = 0.52$ s and $c_2 = -0.11$ s$^2$ from \cite{akhras1981relationship}.

Similar to arterial compartments, the left heart pressure $P_{lh}$ and volume $V_{lh}$ are related by 
\begin{equation}
    P_{lh}-P_{lh,u} = E_{lh}(t)(V_{lh}-V_{un}),
\end{equation}
where $P_{lh,u}=0$ and $V_{lh,u}=10$ are the unstressed pressure and volume in the left heart, and $E_{lh}(t)$ is the time-varying elastance. 
\begin{table}[ht!]
    \centering
    \caption{Dependent variables (volume $V$ (ml), pressure $P$ (mmHg), and flow $Q$ (ml/s) ) for the cardiovascular system and baroreflex control system. The latter includes peripheral vascular resistance $R_{up}$ and $R_{lp}$, left ventricular elastance $E_{lv}$, and heart rate $H$. }
    \begin{tabular}{|l l l c|}
 \hline
    \multicolumn{4}{|c|}{State variables}\\
    \hline\hline
    Symbol & Description & States & Units\\
    $R$ & Resistance & Upper peripheral ($up$) & $\text{mmHg}\cdot \text{s}/\text{ml}$\\
    & & Lower peripheral ($lp$) & $\text{mmHg}\cdot \text{s}/\text{ml}$\\
    
    $E$ & Elastance & Left heart ($lh$) & $\text{mmHg}/\text{ml}$ \\
    & & Diastolic value of left ventricle elastance ($ED$) & $\text{mmHg}/\text{ml}$\\
     
    $P$ & Pressure & Left heart ($lh$) & mmHg\\
    & & Arteries, upper ($au$) & mmHg\\
    & & Arteries, lower ($al$) & mmHg\\
    & & Veins, upper ($vu$) & mmHg\\
    & & Veins, lower ($vl$) & mmHg\\
    & & Mean ($m$) & mmHg\\
      
    $V$ & Volume & Left heart ($lh$) & ml\\
    & & Arteries, upper ($au$) & ml\\
    & & Arteries, lower ($al$) & ml\\
    & & Veins, upper ($vu$) & ml\\
    & & Veins, lower ($vl$) & ml\\
      
    $Q$ & Flow & Atrial valve ($av$) & ml/s\\
    & & Arteries $(a)$& ml/s\\
    & & Upper peripheral $(up)$ & ml/s\\
    & & Lower peripheral $(lp)$ & ml/s\\
    & & Veins ($v$) & ml/s\\
    & & Mitral valve ($mv$) & ml/s\\
      
    $H$ & Heart rate & - & bps \\
    \hline
    
    \end{tabular}
    \label{tab:State table}
\end{table}

\subsection{Head-up tilt (HUT)}

\begin{figure}[ht!]
    \centering
    \includegraphics[scale = .5]{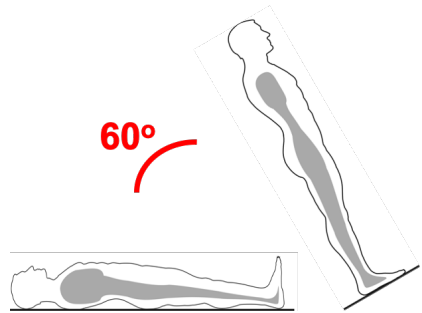}
    \caption{Depiction of a head-up tilt (HUT) test. Patients are tilted, head up, from 0 to 60$^\circ$ over of 7 seconds.}
    \label{fig:HUT}
\end{figure}

During HUT, gravity pools blood from the upper to the lower body affecting the flow between the upper and lower body ($Q_{a}$ and $Q_{v}$). This maneuver is depicted in figure \ref{fig:HUT}. We model this effect by adding a tilt term accounting for the additional force caused by gravitational pooling \cite{williams2014patient}, i.e.
\begin{equation}
    Q_{a} = \frac{P_{au} - P_{al} + P_{tilt}}{R_{a}}, 
    \ \ \ \ \ \
    Q_{v} = \frac{P_{vu} - P_{vl} - P_{tilt}}{R_{v}},
\end{equation}
where
\begin{equation}
    P_{tilt} = \rho g h \sin\Big(\frac{\theta \pi}{180}\Big), \ \ \ \theta \in [0^\circ, \dots, 60^\circ].
\end{equation}

\subsection{Baroreflex model}

The baroreflex (BR) control system maintains homeostasis. Afferent baroreceptor nerves sense changes in the aortic arch and carotid sinus blood pressure (both are included in the compartment representing the upper body arteries).  Signaling in afferent baroreceptor neurons stimulated by blood pressure are integrated into the medulla, from which efferent neurons are activated, modulating signaling along sympathetic and parasympathetic neurons. In the systemic circulation, parasympathetic neurons mainly modulate heart rate, while sympathetic neurons modulate cardiac contractility, peripheral vascular resistance, and vascular elastance. 

\begin{figure}
    \centering
    \includegraphics[scale =.5]{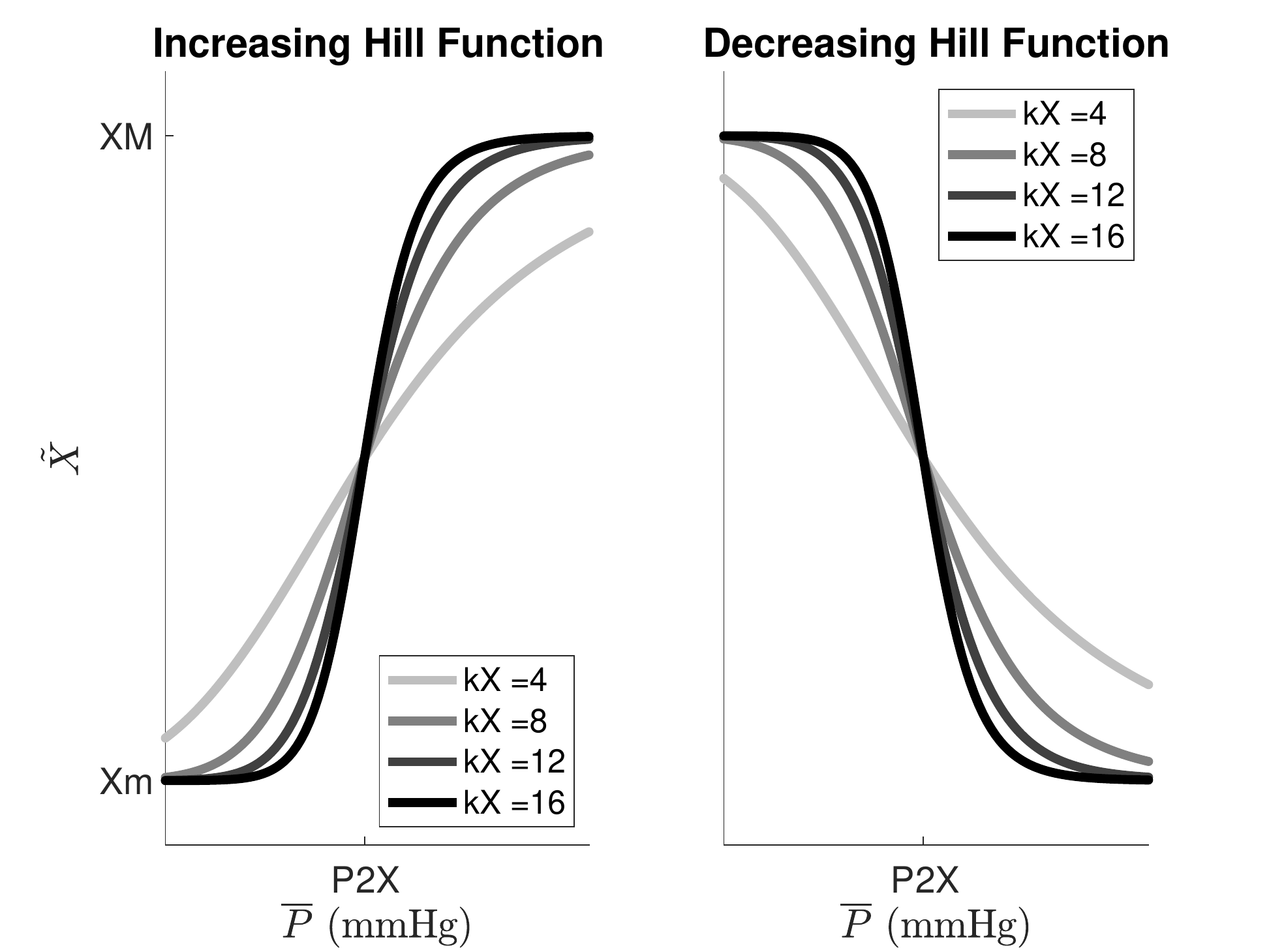}
    \caption{Increasing and decreasing Hill functions for varying Hill-coefficients $k_X$.}
    \label{fig:Hill functions}
\end{figure}

This study includes a simple model directly modulating effector sites in response to changes in blood pressure. Equations for the effector variables $X=\{R_{up},R_{lp}, E_m, H\}$ (listed with units in Table \ref{tab:Hill control table}) are derived under the assumption that each response has a saturation point and a resting value. This assumption motivates the use of first-order kinetic control equations given by
\begin{equation}\label{eq:first order control}
    \frac{dX}{dt} = \frac{-X + \tilde{X}(\bar{P})}{\tau_X},
\end{equation}
where $\tau_X$ is the time-constant for the response $X$ (shorter for effectors stimulated via parasympathetic than via sympathetic neurons). $\bar{P}$ denotes the average BP, computed as  
\begin{equation}
    \frac{d\bar{P}}{dt} = \frac{-\bar{P}+P_{au}}{\tau_P}
\end{equation}
and $\tilde{X}$ is an increasing or decreasing Hill function of the form
\begin{equation}\label{eq:increasing hill}
    \tilde{X} = (X_M-X_m)\frac{\bar{P}^{k_X}}{\bar{P}^{k_X}+P_{2X}^{k_X}} + X_m
\end{equation}
or
\begin{equation}\label{eq:decreasing hill}
    \tilde{X} = (X_M-X_m)\frac{P^{k_X}_{2X}}{\bar{P}^{k_X}+P_{2X}^{k_X}} + X_m
\end{equation}
where $X_M$ is the maximum value of $\tilde{X}$, $X_m$ is the minimum, $P_{2X}$ is the half-saturation value, and $k_X$ is the Hill coefficient. Graphs of increasing and decreasing Hill functions, with varying $k_X$ are shown in Figure \ref{fig:Hill functions}. 

\begin{table}[ht!]
    \centering
    \caption{Quantities controlled by the Baroreflex system using Hill functions. Columns correspond to the quantity being controlled, the symbol of control equation, whether the hill function used is increasing or decreasing, and units.}
    \begin{tabular}{|l l c c|}
    \hline
        Quantity being controlled & Symbol  & Increasing/Decreasing & Units\\
        \hline \hline
         Resistance, upper peripheral & $\tilde{R}_{up}$ & Decreasing & mmHg $\cdot$ s/ml\\
         Resistance, lower peripheral & $\tilde{R}_{lp}$ & Decreasing& mmHg $\cdot$ s/ml\\
         Elastance at end diastole & $\tilde{E}_{D}$ & Increasing & mmHg/ml\\
         Heart Rate & $\tilde{H}$ & Decreasing& bps\\
         \hline
    \end{tabular}
    \label{tab:Hill control table}
\end{table}

\subsection{Heart rate variability (HRV)}

In addition to changes in blood pressure mediated by the baroreflex control system, heart rate data exhibit spontaneous variation, referred to as heart rate variability \cite{Taskforce96}. This is likely due to fluctuations in vagal firing and has proven to be essential for cardiovascular dynamics. While it is well established that HRV is associated with fluctuations in vagal firing setting up a mechanistic model predicting HRV is challenging. To circumvent this, several studies have referred to HRV as  mathematical chaos \cite{goldberger1991normal,shaffer2017overview}. This study examines the importance of including HRV, which we model by  adding ``noise" to heart rate predictions.

We solve the differential equations one cycle at a time, and use current heart rate to determine the length of the next cycle $T = 1/H$. HRV is obtained by adding noise to each cardiac cycle sampled from a uniform distribution, i.e., we let 
\begin{equation}
     T \Big|_{tH0} \leftarrow \frac{1}{H\big|_{tH0}} \cdot \Big(1 + \frac{\textbf{U}[-1,1]}{50} \Big), \label{control noise}
 \end{equation}
 where \textbf{U}$[-1,1]$ is a uniform random distribution from -1 to 1, and $tH0$ denotes the starting time for each heartbeat. We choose to scale the noise by 2\% to approximately match patient data. 
 
\subsection{Model parameters and initial conditions}

Nominal parameter values and initial conditions are deduced from literature and physiological data representing a healthy young female. Below we describe \textit{a priori} calculation of the parameters, which are listed with units in table \ref{tab:Parameters}. This Table include values used to simulate both control and POTS phenotypes.
% JUSTEN ADD PARAMETER VALUES

\subsubsection{Cardiovascular parameters}

\paragraph{Blood volume:} Blood volume is calculated using height and Body Mass Index (BMI). Combining the classic formula for BMI \cite{WHOBMI} by which weight is given by $W = \text{BMI}\big(\frac{h}{100}\big)^2$, with Nadler's equation for blood volume (BV) \cite{nadler1962prediction},to estimate female blood volume (ml) as
$$\text{BV} = 0.4948 \, \text{BMI}^{0.425} \,  h^{1.575} - 1954$$
and male blood volume (ml) as
$$\text{BV} = 0.4709 \, \text{BMI}^{0.425} \, h^{1.575} - 1229$$
where $h$ is height in cm. For our baseline patient we use the average height of women in Denmark, 167.2 cm \cite{ncd2016century}, and a ``healthy" BMI of 22 kg/m$^2$ \cite{WHOBMI}. 

The total blood volume (BV) is distributed between the systemic (containing $\approx$85\%) and pulmonary (containing $\approx$15\%) circulations \cite{Boron}. Within the systemic circulation, at rest, we assume that approximately 15\% is in the arteries and 85\% is in the veins. In the supine position, we assume that 80\% of the blood is in the upper body, while only 20\% is in the lower body \cite{Boron}. To predict circulating blood volume, we differentiate the volume between stressed (circulating) and unstressed volume. Following Beneken and DeWitt \cite{Beneken1967physical}, in the arteries, we assume that 30\% of the volume is stressed, while in the veins we assume that 7.5\% of the total volume is stressed. 

\paragraph{Blood pressure:} The model is parameterized to represent dynamics in a healthy young female with a systolic arterial pressure of 120 mmHg and diastolic arterial pressure of 80 mmHg \cite{lapum2018vital}. Using standard clinical index \cite{demers2020MAP}, we compute the mean pressure as $P_{m} = (2/3)\cdot P_{dia} + (1/3) \cdot P_{sys} \approx 93$ \cite{meaney2000formula}. As blood pressure is typically measured in the arm, which is included in the upper body arteries, we assign these values to the upper arterial compartment. To allow for blood flow from the upper to the lower body arteries, we set the lower body artery pressure to 0.98 times values in the upper body. Since the venous circulation's pulse-pressure is small, we only determine mean values in venous compartments. Using standard literature values \cite{Boron} we assume that the upper body venous pressure is 3 mmHg, again to ensure flow in the correct direction, the lower body venous pressure is $P_{vl} = 1.1 \cdot P_{vu}$. 

Parameters for the lower body venous pressure-volume equation (equation \ref{eqn:Hardy pressure}) are calculated as
\begin{align*}
    V_{Mvl} &= 4 \cdot V_{vlI}\\
    m_{vl} &= \frac{1}{P_{vlI}}\log\Big(\frac{V_{Mvl}}{V_{Mvl}-V_{vlI}}\Big)
\end{align*}
where $V_{vlI}$, $P_{vlI}$ and $V_{Mvl}$ is the nominal volume, pressure and maximal volume for the lower venous compartment ($vl$) respectively. $V_{Mvl}$ is set such that the volume does not saturate at HUT, and $m_{vl}$ is set such that at rest, blood flows from the lower to the upper body veins. 

\paragraph{Elastance:} To calculate nominal Elastance values, we use equation \ref{eqn:compliance pressure eq vol}. Assuming that $P_{un}=0$ and the stressed volume fractions discussed above, we predict elastance for arterial compartments and the upper veins. Due to our use of a non-linear venous pressure-volume equation (equation \ref{eqn:Hardy pressure}), we do not calculate lower venous elastance explicitly. To capture effect of changing the pulse-pressure during HUT this parameter is adjusted following the HUT onset.

\noindent\underline{Left heart end-diastolic and end-systolic elastance:} At the end of diastole, the pressure of the left heart is approximately equal to the venous pressure, and the ventricular volume is maximal, i.e., the nominal (minimal) elastance at diastole can be approximated by
\begin{equation} E_D = P_{vu}/\max(V_{lh}). \end{equation}
Similarly, at the end of systole, the left ventricular pressure is approximately equal to the arterial pressure, and the volume is minimal, giving
the nominal (maximal) elastance at systole 
\begin{equation}\label{eqn:E_S}E_S = P_{au}/\min(V_{lh}).\end{equation}

\paragraph{Blood flows:} In a healthy human cardiac output (CO) is approximately 5 L/min \cite{Boron}. We assume that the total blood volume is circulated in approximately 60 seconds, i.e., the cardiac output CO$\approx BV/60$ ml/s. We assume that all organs above the pelvis, including the gastrointestinal tract, belong to the upper body, while the lower pelvic region and the legs belong to the lower body. With this distinction, we estimate that 80\% of cardiac output travels through the upper peripheral, perfusing the upper body, while 20\% perfuse the lower body \cite{williams1989reference}. Hence we obtain nominal values of $Q_{up} = 0.8$CO, and $Q_{a} = Q_{lp} = Q_{v} = 0.2$CO. 

\paragraph{Resistance:} The atrial and mitral valve resistance are both set to $0.0001$, as we assume the valves do not have significant resistance compared to resistance generated by flow through the vasculature. Therefore, the remaining nominal resistances are calculated using Ohm's law, $R = (P_{i-1} - P_i)/Q$, where $P_{i-1}$ is the pressure in the previous compartment, $P_i$ is the pressure in the destination compartment, and $Q$ is the flow.

\subsubsection{Baroreflex control parameters}

Each control equation has 4 parameters, $\tau_X, X_M,X_m$, and $P_{2X}$. The time-constant, $\tau_X$, reflects the ratio of the speed of the neurological responses and time for the physiological control to occur. We note that the parasympathetic branch ($\tau_H$) operates faster than the sympathetic ($\tau_{R}, \tau_E$), i.e., 
\[
    \tau_H < \tau_{R} = \tau_E
\]
Following our calculation of the base parameters, at rest, we assume that controlled parameters are at equilibrium, i.e., at $P=P_0, \frac{dX}{dt}=0$. Using this assumption, we can estimate $P_{2X}$ as 
\begin{align*}
P_{2X}&=P_0 \Big(\frac{X_M-X_0}{X_0-X_m} \Big), && X = E_D,\\
P_{2X}&=P_0 \Big(\frac{X_0-X_m}{X_M-X_0} \Big), && X = R_{up},R_{lp},H.
\end{align*}
The maximum heart rate ($H_M$) is set to 3.3 bps \cite{nes2013age}, and we assume a minimum heart rate of $H_m=0.3$ bps, representing the smallest sustainable heart rate possible in humans.  We allow peripheral vasculature to dilate to 1.5 times the resting radius and constrict to 0.75 times the resting radius. To relate these measurements to changes in resistance we note that resistance changes in proportion to the fourth power of the radius. Thus we assume that $R_m = 0.2 \cdot R_I$ and $R_M = 3 \cdot R_I$. 

To estimate $E_{DM}$, we refer to the increased potassium levels, which increases cardiac contractility, that has been recorded during exercise in humans \cite{lindinger1995potassium}. From this work, we estimate the extent to which contractility can increase under stress and assume that maximum end-diastolic elastance control ($E_{DM}$) can increase to 125\% of the initial value. In principle, the heart can relax completely by lack of stimulus. We therefore set the minimum end-diastolic elastance control ($E_{Dm}$) to 1\% of the initial value.

\subsubsection{Initial conditions}

Initially, we set the phase of the heart to the end of diastole. Thus $V_{lvI} = 110 - V_{un}$, where 110 is the maximum volume of the left ventricle, and $V_{un}$ is the unstressed volume of the left ventricle, which we assume to be 10 ml \cite{cain2009age}. We also assume a physiologically healthy resting initial heart rate of 1 bps. The remaining initial conditions are set to the calculated nominal values. 

\begin{table}[ht!]
\caption{Table of parameters, their descriptions, values, and units. The character ``$X$" represents the control quantities that can be found in table \ref{tab:Hill control table}. Unit abbreviations are: mmHg - millimeters of mercury, s - seconds, ml - milliliters, bps - beats per second, N.D. - no dimension, ED - end diastole, ES - end systole, coef. - coefficient. The cardiovascular parameters were the same for all phenotypes simulated, while the control parameters differ. Table  \ref{tab:Hill control table} lists values used for each phenotype.}
\label{tab:Parameters}
\begin{tabular}{|l l c c l|}
    \hline
     Symbol & Description &Value & Units & Ref\\
    \hline \hline
    $h$ & Height  &167.2& cm & \cite{ncd2016century}\\
    BMI & Body mass index &22 & kg/m$^2$ & \cite{WHOBMI}\\
    $BV$ & Total blood volume &3887.9&  ml&\cite{du1916clinical,nadler1962prediction}\\
    $V_{un}$ & Unstressed volume  &10& ml &\cite{cain2009age}\\
    $R_{av}$ & Aortic valve resistance & 0.0001&mmHg s/mmHg &\\
    $R_{mv}$ & Mitral valve resistance & 0.0001&mmHg s/mmHg &\\
    $R_{a}$ & Resistance of systemic arteries &0.072& mmHg s/mmHg &\cite{marquis2018practical}\\
    $R_{v}$ & Resistance of systemic veins &0.019& mmHg s/mmHg & \cite{marquis2018practical}\\
    $E_{al}$ & Elastance lower body arteries & 3.1&ml/mmHg &\cite{Boron} \\
    $T_{S}$ & Time-fraction for maximum systole &0.12& s &\cite{akhras1981relationship,janssen2010kinetics,kovacs1985duration}\\
    $T_{D}$ & Time-fraction for minimum diastole & 0.14& s & \cite{akhras1981relationship,janssen2010kinetics,kovacs1985duration}\\
    $E_{ES}$ & ES ventricular elastance & 2&mmHg/ml & Equation \ref{eqn:E_S}\\
    $k_R$  & Resistance Hill coef.  & 23& N.D. &\\
    $k_E$  & Diastolic ventricular elastance Hill coef.  & 7& N.D. &\\
    $k_H$  & Heart rate Hill coef. & 27& N.D. &\\
    $\tau_R$ & Resistance time-constant  &12.5& s & \\
    $\tau_E$ & ED ventricular elastance time-constant  &12.5& s & \\
    $\tau_R$ & Heart rate time-constant &6.25& s & \\
    $\tau_P$ & Mean arterial pressure time-constant  &2.5& s & \\
    $X_{M}$ & Maximum value controlled effector & & &\\
    $X_{m}$ & Minimum value controlled effector & & &\\
    $P_{2X}$ & Half-saturation coef. (pressure) & & mmHg &\\
    \hline
\end{tabular}
\end{table}

\subsection{Signal processing}\label{sig proc methods}

\begin{figure}
    \centering
    \includegraphics[width = \textwidth]{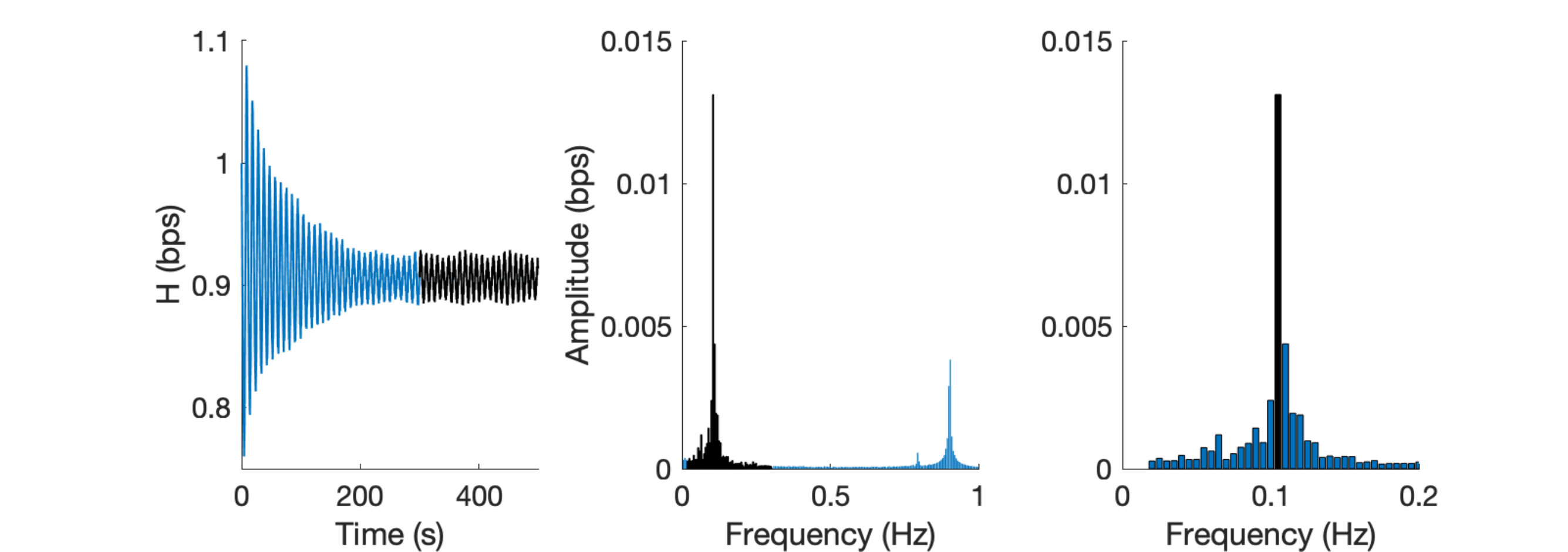}
    \caption{Process of obtaining amplitude of the 0.1 Hz component of a heart rate signal, the same process is used for blood pressure. Starting from the left, first the simulation is ran using a variable step size solver and is interpolated at 100 Hz. The last 200 seconds are recorded (darkened portion). Second, the discrete Fourier transform is applied to obtain amplitude information of the frequencies. We select and examine the range corresponding to the baroreflex, 0.05 - 0.15 Hz (darkened). Third, we find the maximum of the amplitude in this range (darkened).}
    \label{fig:signal processing howto}
\end{figure}

To characterize oscillations seen in the model output, we employ stationary signal processing. This process is illustrated in Figure \ref{fig:signal processing howto}. Once the model has been simulated using MATLAB's \cite{matlab} \verb|ode15s| and there are no transient effects, we interpolate over the data to obtain a time-series that is sampled uniformly at 100 Hz. We then select the last 200 seconds of the $H$ and $P_{au}$ time-series and compute the one-sided power spectrum using MATLAB's FFT algorithm. The design of our model suggests two explainable oscillations: one representing the baroreflex, which operates at approximately 0.1 Hz, and the heart rate, which operates at approximately 1 Hz. As can be seen in Figure \ref{fig:signal processing howto} these two oscillations and their harmonics are the only significant spikes in the frequency domain. To quantify the magnitude of oscillations caused by our control equations, we record the power and phase of the maximum amplitude peak in the $\approx 0.1$ Hz frequency range. This process is applied both for the model at rest and computed again for the HUT portion.

\subsection{Emergence of low-frequency oscillations}\label{sec:parameter analysis}

To capture the emergence of low-frequency $\approx$0.1 Hz oscillations we first conducted a parameter sweep changing all relevant parameters over their physiological range. Specific emphasis is on parameters in equations facilitating baroreflex control (Equations \ref{eq:first order control}-\ref{eq:decreasing hill}). This analysis is done in two steps, first detecting what parameters impact the dynamic behavior and second conducting a detailed analysis varying critical parameters impacting the dynamic response. In addition, to detecting what parameters cause the model to change behavior, we also investigate how to set parameters to capture oscillations at the $\approx0.1$Hz frequency range. 

Pseudo-code for this analysis is included in Table \ref{tab:2 dim pseudo}. The model is solved initially for 250 seconds. After 250 seconds the solution is examined to verify that the solution has reached steady state for 200 seconds. To verify steady state behavior the previous 200 seconds are split in half and maximum and minimum heart rate ($H$) and upper arterial blood pressure ($P_{au}$) are calculated for each half and the relative difference between values for each half is computed and compared to a threshold ($\alpha$). The halves are then interpolated at 100 Hz and Fourier power spectra are computed for heart rate and blood pressure for each half. The maximum $\approx 0.1$ Hz power value is recorded for each half and the relative difference between the power of the halves are computed and compared to a threshold ($\beta$). If all of these relative differences are less than their respective thresholds the model is said to be a steady state and the power spectra of the last 200 seconds is computed and recorded. If at least one of the relative differences is above the threshold the model is solved for 20 additional seconds and checked for steady state behaviour again.

This approach allows for an automated way to explore the parameter space. Doing so we can observe Hopf bifurcations with respect to the low-frequency oscillations and can observe the effects of parameters on the strength of oscillations.

\begin{table}[!ht]
    \centering
    \begin{tabular}{|l l l|}
    \hline
    \multicolumn{3}{|c|}{Two-dimensional parameter analysis pseudo-code}\\
    \hline \hline 
    \multicolumn{3}{|l|}{\textbf{Input}: Parameter values } \\
    \multicolumn{3}{|l|}{\textbf{Output:} Amplitude and frequency of $\approx$ 0.1 Hz response, $H$ \& $P_{au}$}\\
        \multicolumn{1}{|r}{$\alpha$} &$\leftarrow 0.01$  &\\
        \multicolumn{1}{|r}{$\beta$} & $\leftarrow 0.1$  & \\
        \multicolumn{1}{|r}{$T_2$} & $\leftarrow 250$  & \\
        \multicolumn{3}{|l|}{\textbf{While}:  $t\leq T_2$ } \\
        & \multicolumn{2}{l|}{Run model for one heartbeat, $t_H \leftarrow$ $t + $ duration of heartbeat}\\
         & \textbf{IF}: $t_H \geq T_2$ & \\
         & & $T_0 \leftarrow T_2 - 200$\\
    & & $T_1 \leftarrow \frac{T0+T2}{2}$\\
    & &  $C_1 \leftarrow  \Big|\frac{\max \big(H([T_0,T_1])\big) - \max \big(H([T_1,T_2]) \big)}{\max \big( H([T_1,T_2]) \big)}\Big| \leq \alpha$\\
    & & $C_2 \leftarrow  \Big|\frac{\min \big(H([T_0,T_1])\big) - \min \big(H([T_1,T_2]) \big)}{\min \big( H([T_1,T_2]) \big)}\Big| \leq \alpha$\\
    & & $C_3 \leftarrow  \Big|\frac{\max \big(P_{au}([T_0,T_1])\big) - \max \big(P_{au}([T_1,T_2]) \big)}{\max \big( P_{au}([T_1,T_2]) \big)}\Big| \leq \alpha$\\
    & & $C_4 \leftarrow  \Big|\frac{\min \big(P_{au}([T_0,T_1])\big) - \min \big(P_{au}([T_1,T_2]) \big)}{\min \big( P_{au}([T_1,T_2]) \big)}\Big| \leq \alpha$\\
    & & Interpolate $H$,$P_{au}$ at 100 Hz to obtain $\tilde{H}$,$\tilde{P}_{au}$\\
    & &  Compute $A_{H1}$, $A_{H2}, A_{P1}, A_{P2}$\\
    & & \textbf{IF}: $A_{H2} > 0$\\
    & & \qquad $C_5 \leftarrow \frac{A_{H1}-A_{H2}}{A_{H2}} \leq \beta$\\
    & & \textbf{ELSEIF}: $A_{H1} == A_{H2}$, $C_5 \leftarrow 1$\\
    & & \textbf{ELSE}: $C_5 \leftarrow 0$\\
    & & \textbf{END}\\
    & & \textbf{IF}: $A_{P2} > 0$\\
    & & \qquad $C_6 \leftarrow \frac{A_{P1}-A_{P2}}{A_{P2}} \leq \beta$\\
    & & \textbf{ELSEIF}: $A_{P1} == A_{P2}$, $C_6 \leftarrow 1$\\
    & & \textbf{ELSE}: $C_6 \leftarrow 0$\\
    & & \textbf{END}\\
    & & \\
    & & \textbf{IF}: $\min(C_1,C_2,C_3,C_4,C_5,C_6) == 0$\\
    & & \qquad $T_2 \leftarrow T_2+20$\\
    & & \textbf{END}\\
    & \textbf{END} &\\
     & $t \leftarrow t_H$ & \\
    \textbf{END} & & \\
    \multicolumn{3}{|l|}{Calculate and record metrics} \\
    \multicolumn{3}{|l|}{\textbf{END CODE}}\\
    \hline
    
    \end{tabular}
    \caption{Pseudo code for two-dimensional parameter analysis. $H([T_i,T_k])$ represents the value of heart rate ($H$) between time of $T_i$ and $T_k$, similarly for blood pressure $P_{au}$. $A_{H1}$ denotes the amplitude of the  0.1 Hz component of the $H$ signal during $[T_0,T_1]$ as is explained in methods section \ref{sig proc methods}.  Similarly, $A_{P1}$ for $P_{au}$. $A_{H2}$ and $A_{P2}$ are denote the amplitude of the approximately 0.1 Hz component during $[T_1,T_2]$. Analysis is conducted for both rest and head-up tilt sections.}
    
    \label{tab:2 dim pseudo}
\end{table}

\clearpage
%Rest:
%Control:  kH,kR,kE, p2X
%POTS:     kH,kR,kE, p2X

%HUT: 
%Control (decreasing upper body arterial compliance - volume distribution) - ref ca paper 2005.
%POTS (increased):     
%kH,kR,kE - hyperadrenergic => do not get sustained osc after HUT smaller than at rest

%POTS (increased):     p2H (tachycardia) [ XM - Xm ]
%neuropathic (lower vascular beds): [Values from hyperadrenergic but decrease XlbM-Xlbm]

\subsection{POTS and its phenotypes}

We model POTS and its phenotypes suggested by \cite{mar2020postural} by adjusting parameters to reflect the hypothesised physiology. The values of the parameters before and after HUT can be seen in Table \ref{table:phenotype params}.

{\em The hyperadrenergic POTS} is characterized by high levels of circulating norepinephrine during postural change, which allows the sympathetic nervous system to be more responsive to changes in blood pressure. To simulate this, at the onset of HUT we further increase parameters associated with sympathetic response including $P_{2H}$ and $k_H,k_E,k_R$.
%JUSTEN "BASE AND PHENOTYPES"

{\em Neuropathic POTS} is caused by partial neuropathy of the lower body vasculature, which causes abnormal pooling of blood in the lower extremities. To simulate this, we decrease the control for the lower body resistance by reducing $ R_{lbM}$ and $R_{lbm}$ after HUT.
% JUSTEN WHY ONLY DURING HUT - I THINK THAT SHOULD BE GENERAL

{\em The hypovolemic POTS} is obtained by decreasing the total blood volume. In isolation, this phenotype does not compromise the baroreflex, and should therefore not be specified as an individual phenotype. However, we do acknowledge that a large number of patients with severe POTS side-effects are young skinny females. However, rather than modeling low blood volume as a phenotype we investigate how the two phenotypes are represented in normotensive and hypervolumic individuals.  

To allow for a smooth transition between parameter values that allows for a 10-second delay, we translate the change in parameters mimicking incorporating a delayed onset, i.e.; 
\begin{equation}
    x = \begin{cases} 
      x_0 & t< t_{HUT}+10 \\
      (x_1 - x_0)\frac{(t-t_{HUT}-10)^8}{(t-t_{HUT}-10)^8 + 5^8} + x_0 & t\geq t_{HUT}+10
   \end{cases}
\end{equation}
where $x$ is the parameter being changed after HUT, $x_0$ is the value of the parameter during rest, $x_1$ is the value of the parameter that is being transitioned to and $t_{HUT}$ is the time of the start of the HUT.

\begin{table}[ht!]
\centering
\begin{tabular}{|l|c|c|c|} 
 \hline 
 Phenotype & Parameter & Value before tilt &Value after tilt\\
 \hline \hline
Control & $C_{au}$ & 1.15 ml/mmHg & 0.69 ml/mmHg\\
& $P_{2H}$& 89 mmHg & 87 mmHg\\
Hyperadrenergic & $k_H$ &27 N.D.&47 N.D.\\
& $k_R$ & 23 N.D. &40 N.D.\\
&$k_E$ & 7 N.D.&12 N.D.\\
&$P_{2H}$&88.7 mmHg &89.6 mmHg\\
& $C_{au}$ & 1.15 ml/mmHg & 0.69 ml/mmHg\\
Neuropathic & $R_{lpM}$ & 4.5 mmHg $\cdot$ s/ml&2.9 mmHg $\cdot$ s/ml\\
& $R_{lpm}$&0.30 mmHg $\cdot$ s/ml&0.19 mmHg $\cdot$ s/ml\\
&$C_{au}$ & 1.15 ml/mmHg & 0.69 ml/mmHg\\
 \hline
\end{tabular}
\caption{Values of selected parameters before and after head-up tilt for phenotype simulations. Parameters are as follows: $C_{au}$ - upper arterial compliance, $P_{2H}$ - half saturation value for heart rate control, $k_H$ - Hill coefficient for heart rate control, $k_R$ - Hill coefficient for resistance control, $k_E$ - Hill coefficient for left heart end diastolic elastance control, $R_{lpM}$ - maximum value of resistance control, $R_{lpm}$ - minimum value of resistance control. }
\label{table:phenotype params}
\end{table}

\section{Results}

Results demonstrate the emergence of low-frequency oscillations at rest and during HUT and how the phenotypes proposed by Mar and Raj \cite{mar2020postural} can be simulated.

\subsection{Low-frequency oscillations}

Our model, shown in Figure \ref{fig:System}, can generate $\approx$0.1 Hz heart rate ($H$) and blood pressure ($P_{au}$) oscillations observed in patient data \cite{Geddes2020Osc}. The amplitude and frequency of the oscillations can be modulated by varying the model parameters in the baroreflex control equations (Equations \ref{eq:first order control}-\ref{eq:decreasing hill}), including  the maximum $X_{M}$ and minimum $X_{m}$ response, the time-constants $\tau_X$, the half-saturation values $P_{2X}$, and the Hill-coefficients $k_X$, $X=H,R,E$. 

\paragraph{Oscillation frequency} is primarily determined by time-constants ($\tau_X$) differentiating the parasympathetic and sympathetic control. Efferent responses mediated by the parasympathetic system are significantly faster than those transmitted via  the sympathetic system \cite{Boron}, i.e., $\tau_H\ll\tau_R=\tau_E$. The 0.1 Hz frequency was achieved using time-constants reported in Table \ref{tab:Parameters}. We observe that the Hill coefficients ($k_X$) also have an effect on frequency but to a lesser extent than $\tau_X$.
%JG:ADD SOMETHING ON OTHER PARAMETERS IMPACTING FREQUENCY? - Answered

\paragraph{Oscillation amplitude} can be modulated by changing $k_X, P_{2X}$ and $X_M-X_m$, $X=H,R,E$. We studied the effect of varying all parameters over their physiological range. Results (summarized in Table \ref{table:param effects}) show that $k_X, X=H,R,E$ impact the oscillation amplitude, with $k_H$ being more influential.  Increasing the Hill-coefficients $k_X$ increases the sensitivity of the baroreflex control. A larger value of $k_X$ gives a steeper set-point function (shown in Figure \ref{fig:Hill functions}), i.e., the change in pressure needed to generate a given response decreases. Shifting the Hill function by changing the half saturation value ($P_{2X}$) can cause the operating regime to change to a steeper portion of the Hill function. This shift causes has the same effect as increasing $k_X$ but has a much smaller effect on oscillation amplitude.
% JG: ADD SOMETHING ON HALF_SATURATION VALUES AS WE MODULATE THESE IN THE CELL PAPER - Answered, reword?

\begin{figure}[ht!]
     \centering
     \begin{subfigure}[b]{\textwidth}
         \centering
         \includegraphics[width=\textwidth]{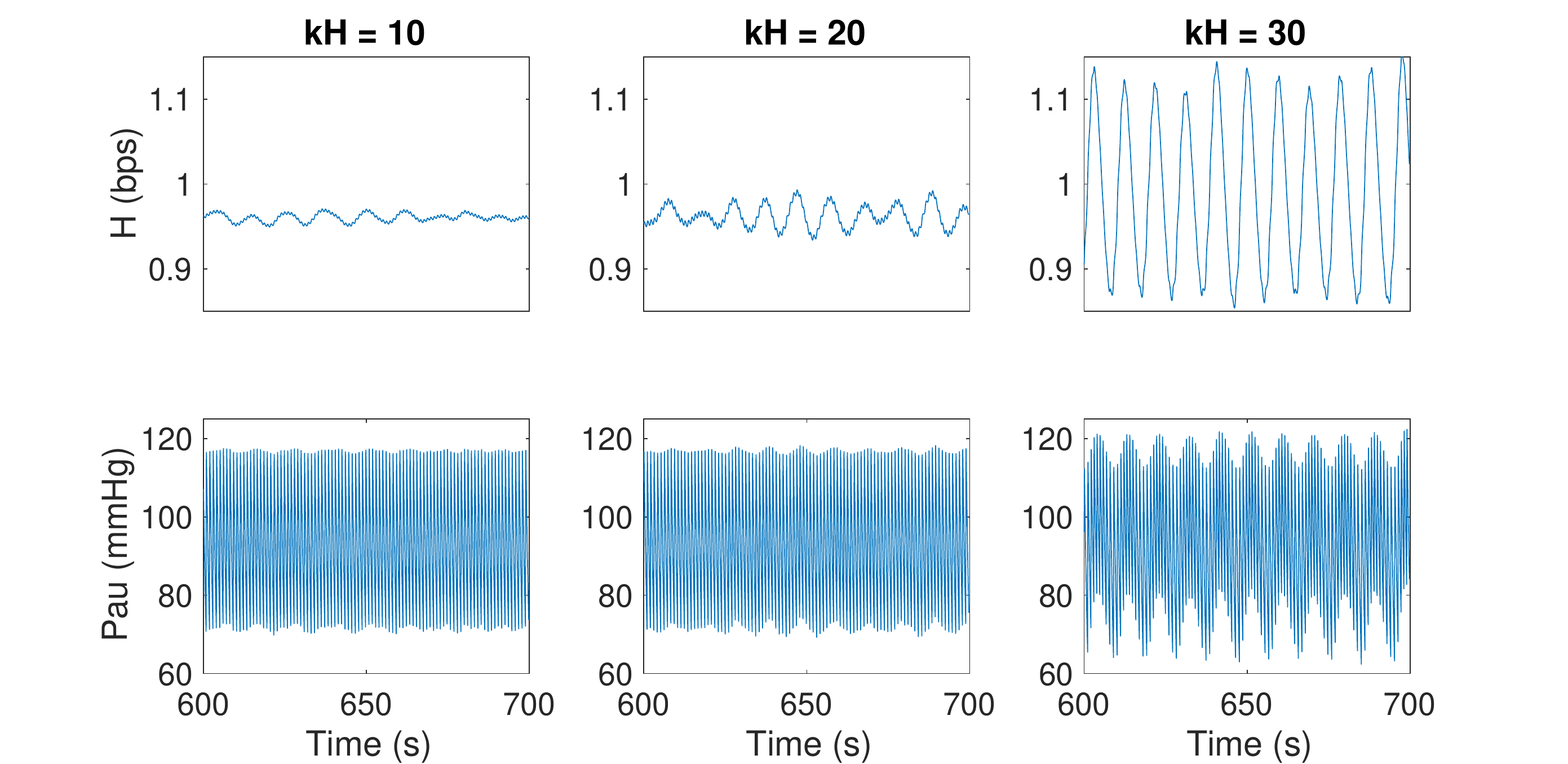}
         \label{fig:5_kH1D_timeseries}
     \end{subfigure}
     \newline
     \begin{subfigure}[b]{\textwidth}
         \centering
         \includegraphics[width=\textwidth]{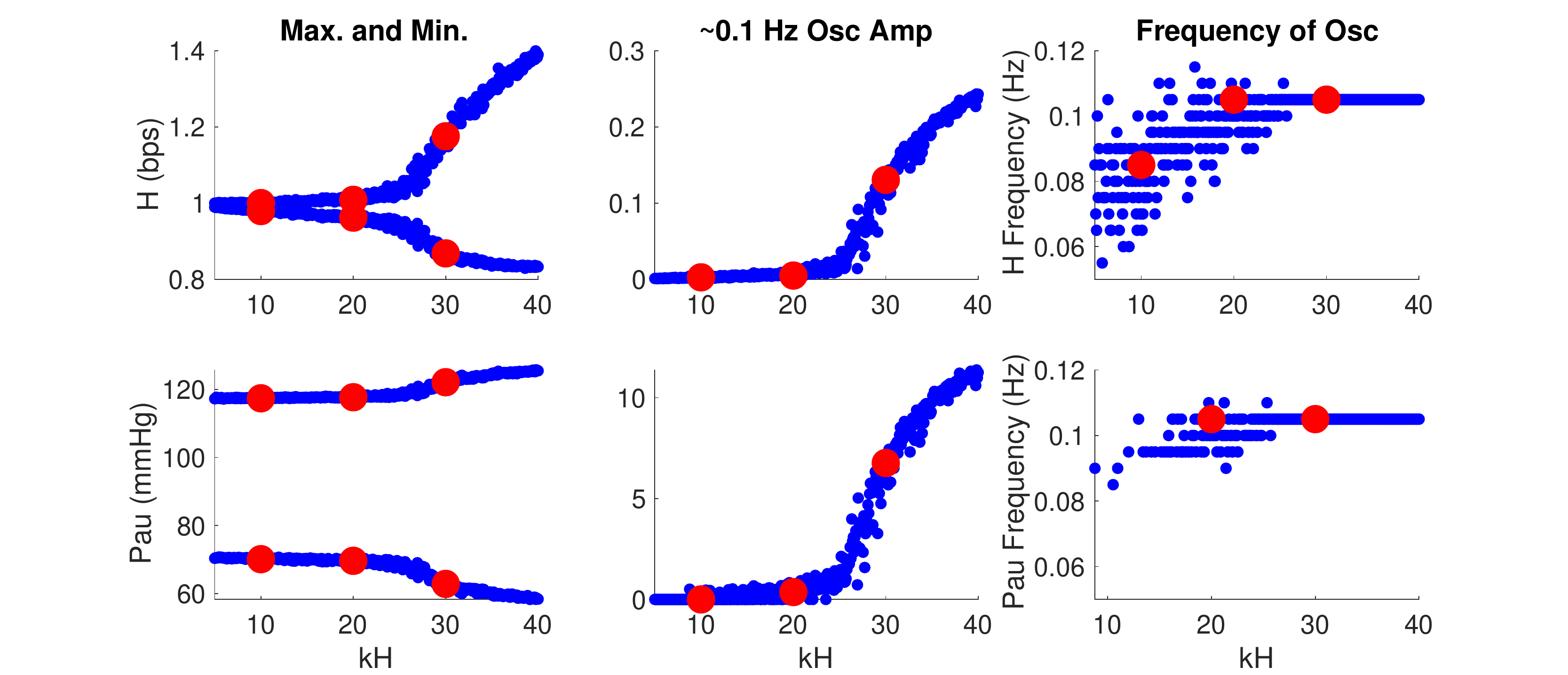}
         \label{fig:5_kH1D_osc_strength}
     \end{subfigure}
      \caption{(A) From left to right: heart rate (H, top) and upper arterial pressure ($P_{au}$, bottom) predictions for $k_H = 10,20,$ and $30$. (B) From left to right: maximum and minimum values of predictions for varying values of $k_H$, the amplitude of the $0.1$ Hz region response, and frequency of oscillations. Enlarged red dotes show denote measurements corresponding to $k_H = 10,20,30$.}
    \label{fig:2by4 osc vs no osc}
\end{figure}

Figure \ref{fig:2by4 osc vs no osc}A (top panels) shows HR and BP dynamics in response to increasing $k_H$. We depict results changing $k_H$ as the most influential parameter, but similar results (not shown) are obtained when increasing $k_R$ and $k_E$, controlled by the sympathetic system. For $k_H< 6$, the system does not oscillate, at $k_H \approx 6$, oscillations emerge, and their amplitude increases with increasing values of $k_H$.  Figure \ref{fig:2by4 osc vs no osc}B depicts the change in amplitude and frequency as a function of $k_H$. Changing $k_X, X = H,R,E$ impacts the oscillation amplitude more than frequency. The frequency almost  doubles (it changes from ~0.6Hz to 1.1 Hz) while the HR oscillation amplitude increases from $\approx$0 to 0.3. The frequency diagrams in Figure \ref{fig:2by4 osc vs no osc}B (right column) have two characteristic features, a broad distribution (vertical spread) and horizontal stripes with white spacing. The former results from noise, added to heart rate, to account for heart rate variability, and the latter by the frequency resolution. The model is solved with a time step of 0.01 s, with the Fourier transform calculated over a 200 s interval, giving a frequency resolution of 0.005 Hz.

\begin{figure}[ht!]
     \centering
     \begin{subfigure}[b]{0.49\textwidth}
         \centering
         \includegraphics[width=\textwidth]{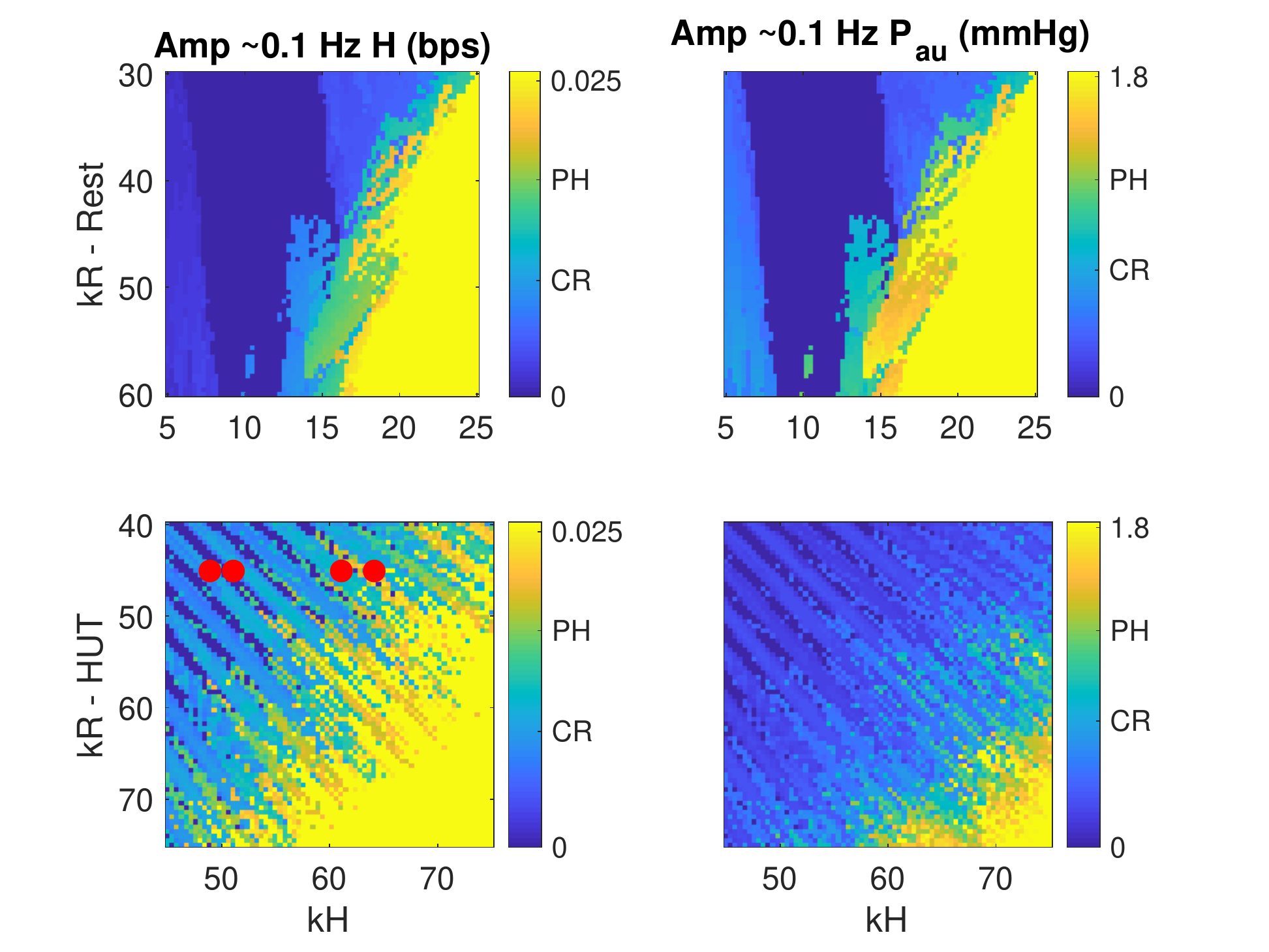}
         \caption{$k_R$ vs $k_H$}
         \label{fig:kRkH sub1}
     \end{subfigure}
     \begin{subfigure}[b]{0.49\textwidth}
         \centering
         \includegraphics[width=\textwidth]{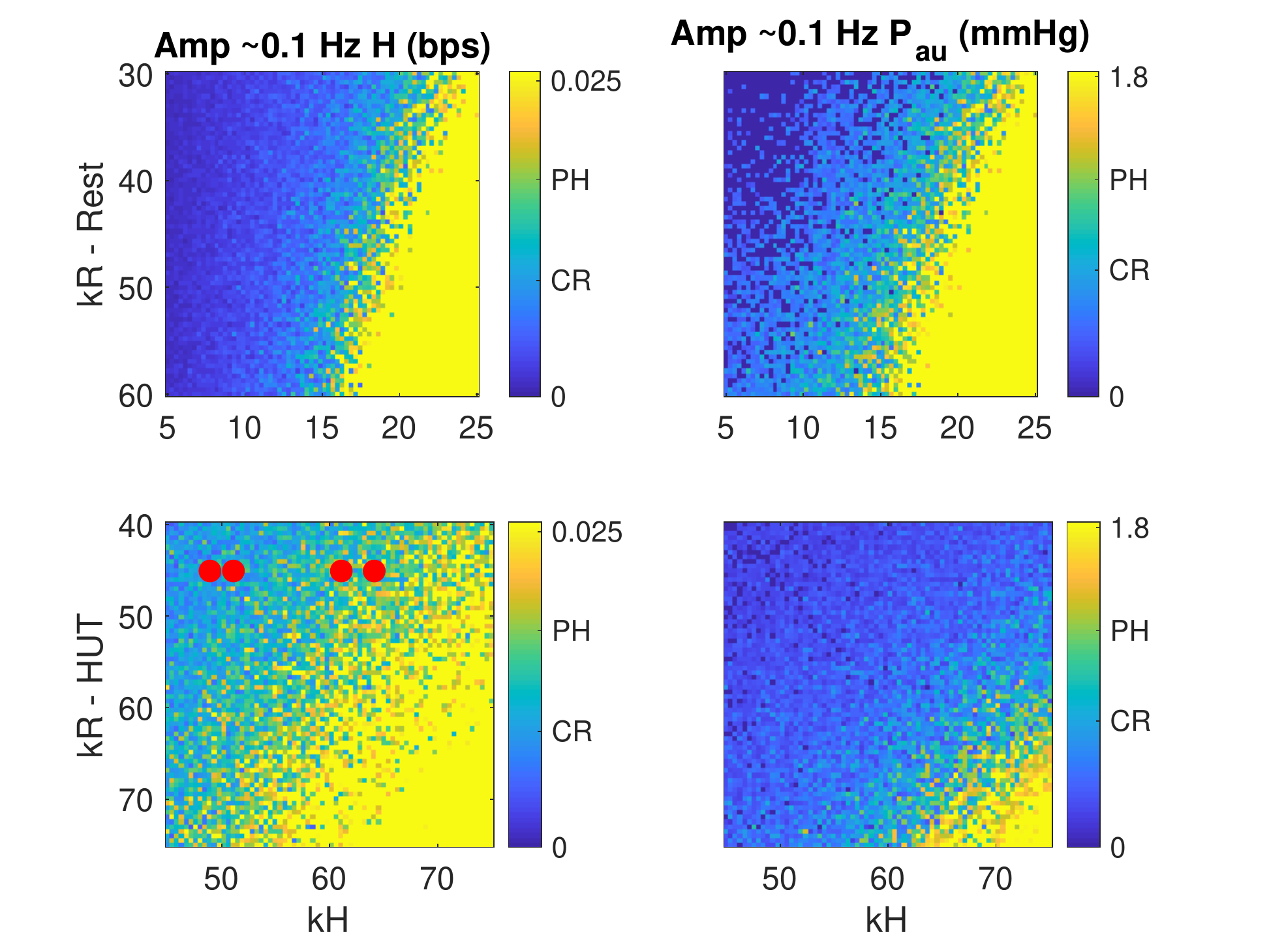}
         \caption{$k_R$ vs $k_H$ with 2\% noise.}
         \label{fig:kRkH noise1}
     \end{subfigure}
     \newline
     \begin{subfigure}[b]{0.49\textwidth}
         \centering
         \includegraphics[width=\textwidth]{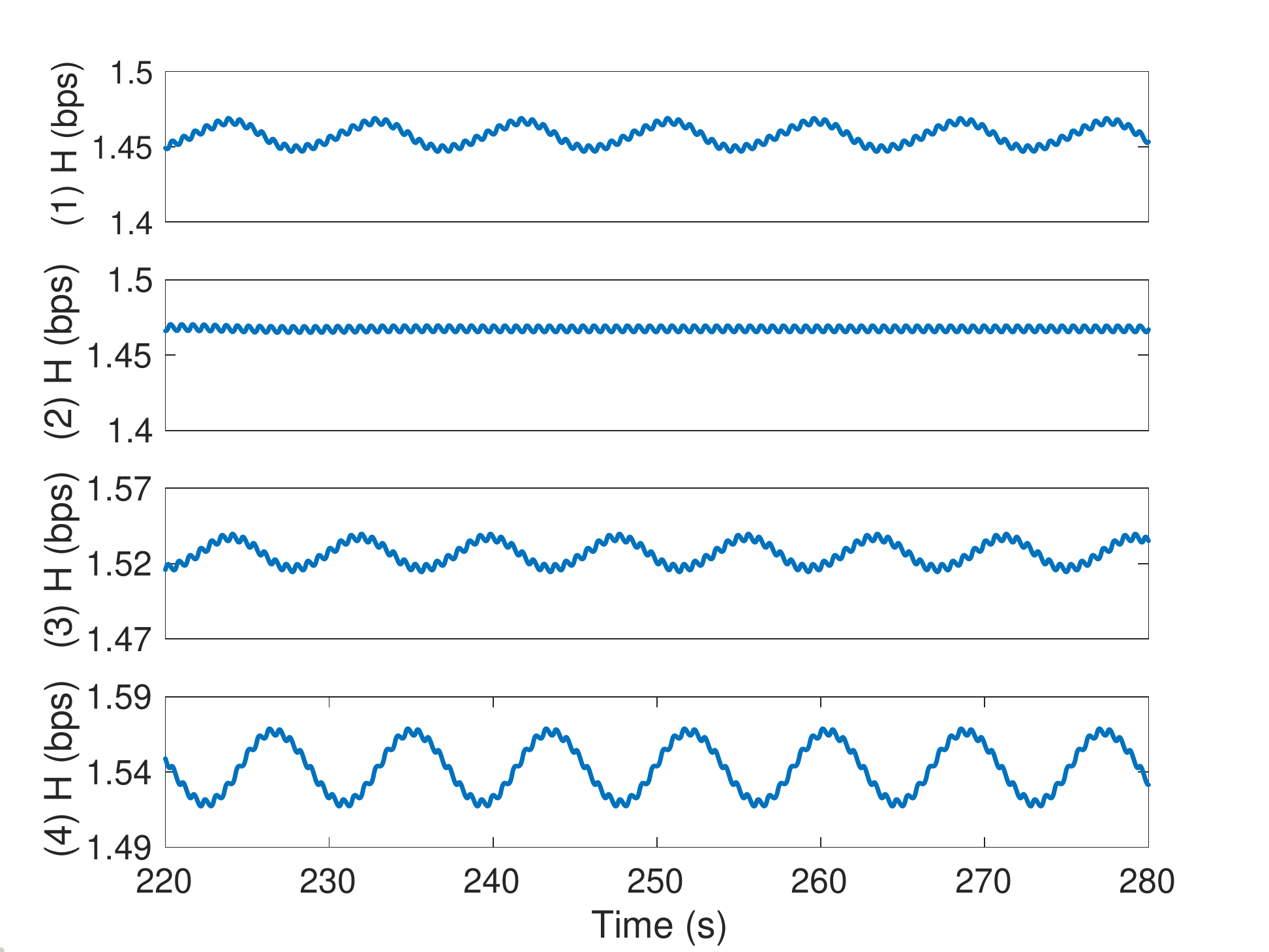}
         \caption{HUT time-series for 4 red dots marked\\ in sub-figure \ref{fig:kRkH sub1}}
         \label{fig:kRkH sub time}
     \end{subfigure}
     \begin{subfigure}[b]{0.49\textwidth}
         \centering
         \includegraphics[width=\textwidth]{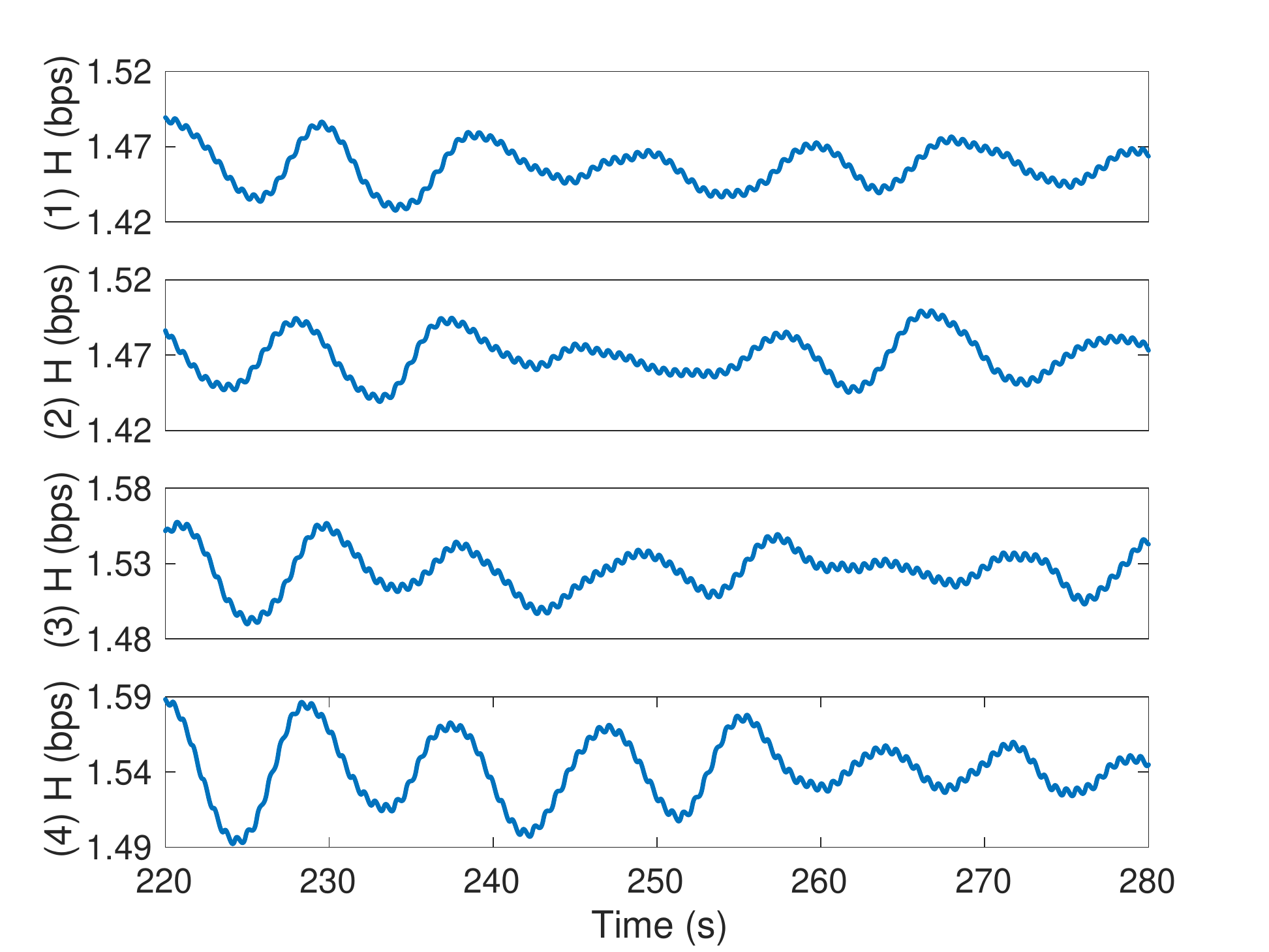}
         \caption{HUT time-series for 4 red dots marked\\ in sub-figure \ref{fig:kRkH noise1}}
         \label{fig:kRkH noise time}
     \end{subfigure}
     \caption{Two-dimensional parameter analysis of $k_R$ vs. $k_H$. (A) Amplitudes of peak heart rate ($H$) oscillation (left) and peak upper arterial blood pressure ($P_{au}$) oscillation (right) at the $\approx 0.1$ Hz frequency band for values of $k_R$ and $k_H$ at rest (top) and head-up tilt (HUT, bottom). (B) The same information as (A) but with 2 \% noise. Average measurements from data \cite{Geddes2020Osc} are marked for Control patients at rest (CR), and POTS patients during head-up tilt (PH). Note that the physiologically possible oscillations correspond to the green regions. (C) Heart rate predictions during HUT for red dots on lower panels of (A); $i^{th}$ panel from top corresponds to $i^{th}$ dot from the left in (A). (D) Similar information as (C) but pertaining to (B).}
     \label{fig:kRkH}
\end{figure}

As noted above, $k_H$ has the most significant impact on the system dynamics.  Both the sympathetic and parasympathetic systems control heart rate, but as noted in the introduction, POTS may be the result of the expression of specific agonistic antibodies binding to $\beta_1$ and $\beta_2$ receptors \cite{mar2020postural}. Since these are found on pacemaker cell modulating heart rate and smooth muscle cells in the vasculature, we study the response to changing $k_H$ and $k_R$. Results shown in Figures \ref{fig:kRkH}A and \ref{fig:kRkH}B reveal that increasing either $k_H$ or $k_R$ increases the amplitude of oscillations. This result agrees with the hypothesis that POTS patients have a more sensitive control system. 

The other model parameters also change the dynamic behavior - but not as significant as changes in $k_X$ (specifically $k_H$). In general, the half-saturation value offsets the control at different pressure levels but does not change the sensitivity, as the slope of the sigmoidal curve remains the same. Changing the range $\Delta X = X_{M}-X_{m}$ changes the width and steepness of the curve;  the latter does have some effect on sensitivity, but it is not as significant as the effect observed when increasing $k_X$.  Table \ref{table:param effects} lists the effects of changing each parameter on HR and BP. 

\begin{table}[!htbp]
\centering
\begin{tabular}{|c|l|} 
 \hline 
 Increased Parameter & Effect\\
 \hline \hline
 $k_H$ & Increases $H$ \& $P_{au}$ oscillation amplitude \\
 $k_R$ & Increases $H$ \& $P_{au}$ oscillation amplitude \\
 $k_E$ & Increases $H$ \& $P_{au}$ oscillation amplitude \\
 $P_{2H}$ & Increases $H$, increases diastolic $P_{au}$\\
 $P_{2R}$ & Decreases $H$, increases $P_{au}$\\
 $P_{2E}$ & Decreases $H$,  increases $H$ \& $P_{au}$ oscillation amplitude \\
 & Increases $P_{au}$ pulse-pressure \\
 $H_M-H_m$ &  Increases $H$ \& $P_{au}$ oscillation amplitude\\
 $R_M-R_m$ &  Increases $H$ \& $P_{au}$ oscillation amplitude\\
 $E_{DM} - E_{Dm}$ & Increases $H$ \& $P_{au}$ oscillation amplitude \\
 
 \hline
\end{tabular}
\caption{Effects on heart rate ($H$) and upper arterial blood pressure ($P_{au}$) when increasing stated parameter.}
\label{table:param effects}
\end{table}

\begin{figure}[ht!]
     \centering
     \begin{subfigure}[b]{0.49\textwidth}
         \centering
         \includegraphics[width=\textwidth]{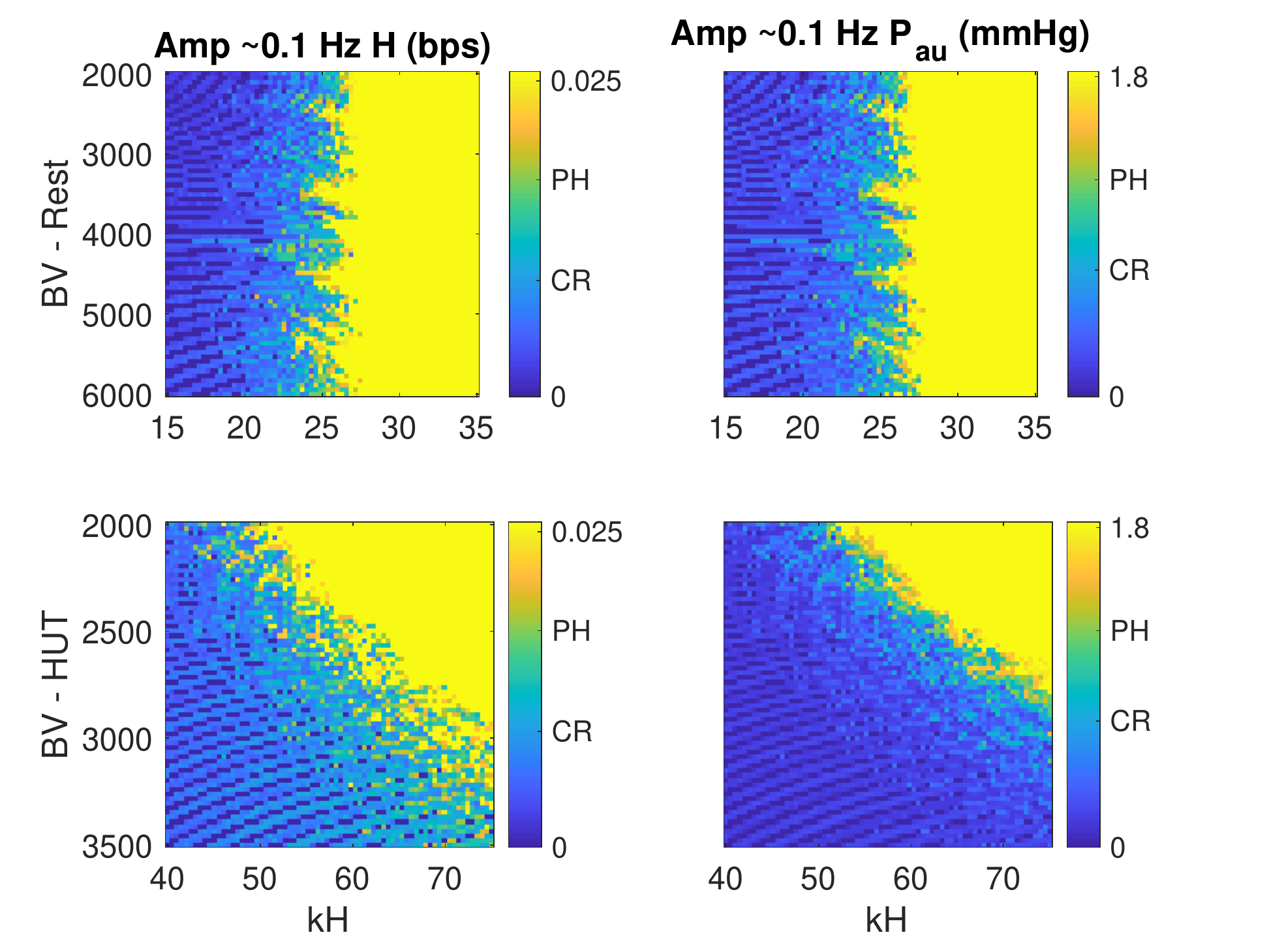}
         \caption{BV vs $k_H$}
         \label{fig:BVkH sub}
     \end{subfigure}
     \begin{subfigure}[b]{0.49\textwidth}
         \centering
         \includegraphics[width=\textwidth]{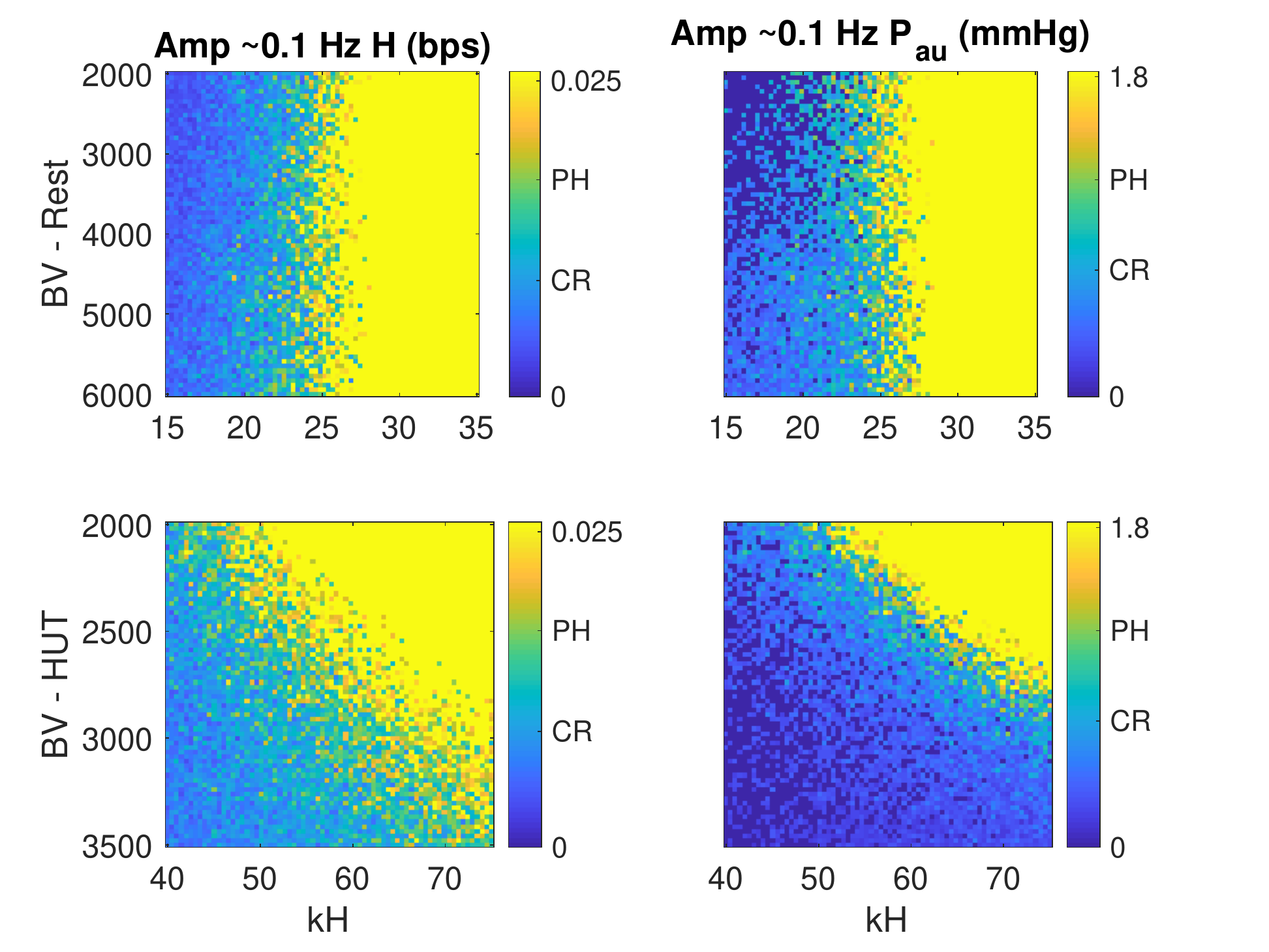}
         \caption{BV vs $k_H$ with 2\% noise.}
         \label{fig:BVkH noise}
     \end{subfigure}
     \caption{Two-dimensional parameter analysis of blood volume (BV) vs $k_H$. (A) Amplitudes of peak heart rate ($H$) oscillation (left) and peak upper arterial blood pressure ($P_{au}$) oscillation (right) at the $\approx$ 0.1 Hz frequency band for values of BV and $k_H$ at rest (top) and head-up tilt (HUT, bottom). (B) The same information as (A) but with 2 \% noise. Average measurements from data \cite{Geddes2020Osc} are marked for Control patients at rest (CR), and POTS patients during head-up tilt (PH).}
     \label{fig:BVkH}
\end{figure}

\subsection{Head-up tilt (HUT)}

\begin{figure}[ht!]
    \centering
    \includegraphics[scale = 0.6]{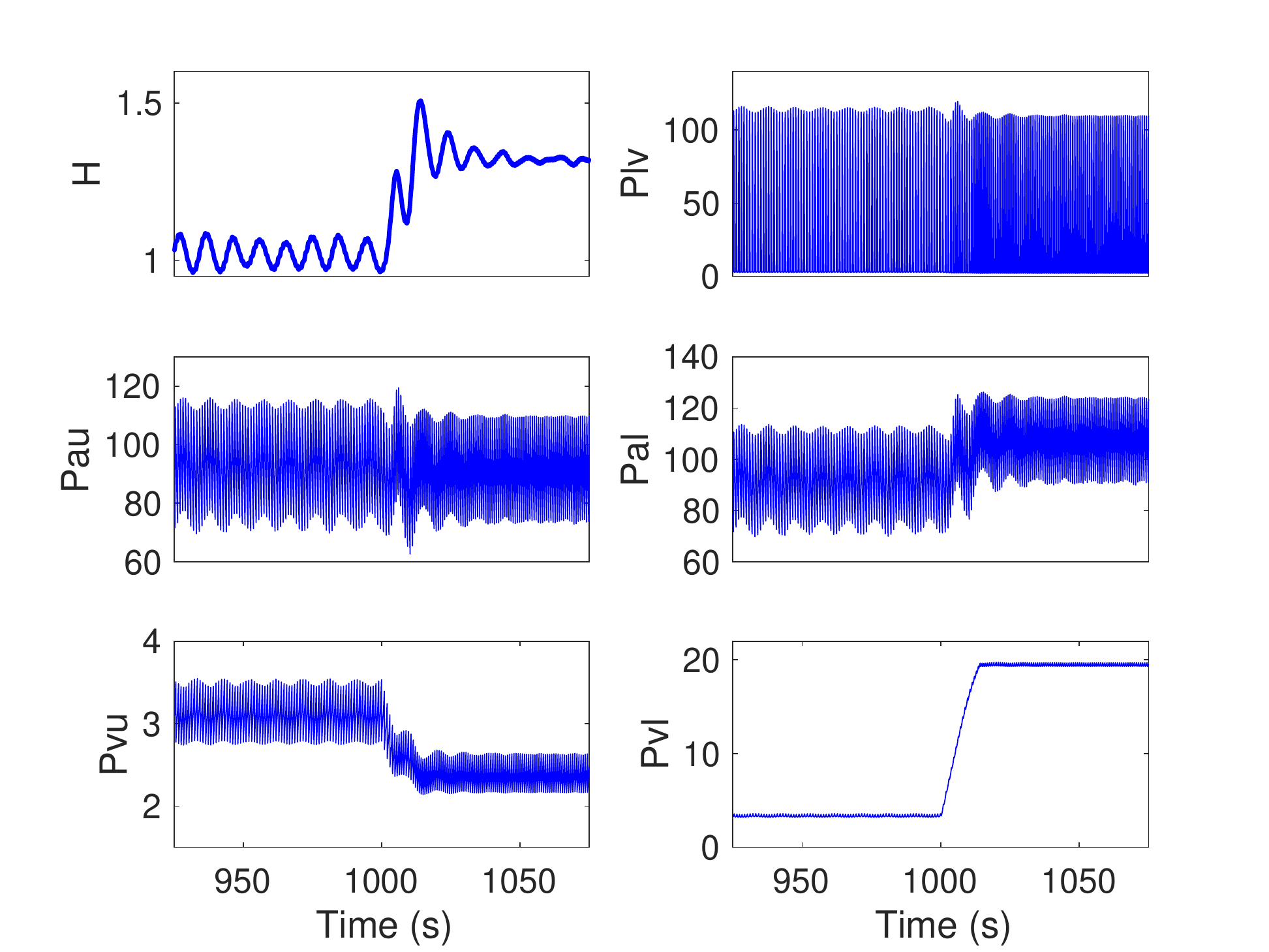}
    \caption{Results of simulation with HUT at $t=1000$. Row 1: heart rate, $H$  (bps), left ventricle pressure, $P_{lv}$ (mmHg) row 2: upper arterial pressure, $P_{au}$ (mmHg), lower arterial pressure, $P_{al}$ (mmHg) row 3: upper venous pressure, $P_{vu}$ (mmHg), lower venous pressure, $P_{vl}$ (mmHg).}
    \label{fig:Compartment Response}
\end{figure}

During HUT (shown in Figure \ref{fig:HUT}), gravity pools blood from the upper to the lower body, stimulating the autonomic nervous system. The result is a shift in blood volume and pressure, increasing in compartments below the center of gravity and decreasing in compartments above. In our model, the upper body compartments are centered around the carotid baroreceptors, while the lower body compartments are centered in the lower part of the torso. Representative model blood pressure predictions in all compartments are shown in  Figure \ref{fig:Compartment Response}. We note that after HUT, the pressure in the lower compartments increases while pressure in the upper compartments decreases. These simulations were generated with $k_H = 27$, which causes the system to oscillate at rest and after HUT. Without changing parameters, oscillations dampen after HUT as a result of volume redistribution.

Similar to rest, control parameters impact predicted dynamics, and $k_H$ remains the most influential parameter. Figures \ref{fig:kRkH}A (bottom row) shows oscillation amplitude as a function of $k_R$ and $k_H$ without noise. For these simulations,  the "non-oscillatory" region appears striped, indicating bands of oscillations alternating with no oscillations. Figure \ref{fig:kRkH}C shows selected time-series predictions for parameter values marked on Figure \ref{fig:kRkH}A. We note that in the non-oscillatory region, it is possible to increase $k_H$ and eliminate oscillations. These stripes are a result of the on-off behavior of emerging low-frequency oscillations. Mathematically, this  behavior is common; as we change $k_R$ or $k_H$ the system undergoes repeated Hopf bifurcations. However, physiologically, small changes in a parameter have not been reported to affect the frequency response significantly.  By adding noise mimicking heart rate variability (HRV) to the model, this behavior disappears (the striped pattern disappears, see Figure \ref{fig:kRkH}B), suggesting that the presence of HRV stabilizes the system response as can be seen in Figures \ref{fig:kRkH}B and \ref{fig:kRkH}D. 

\subsection{POTS phenotypes}

Previous studies \cite{mar2020postural,fedorowski2019postural} suggest that POTS patients can be separated into neuropathic, hyperadrenergic, and hypovolemic phenotypes. This section discusses how each of these can be represented in our model. Note, we do not have blood pressure or heart rate data annotated with specific phenotypes. Therefore the results presented here depict qualitative rather than quantitative.

The phenotype encoding is based on the assumption that the cardiovascular system of POTS patients changes in response to a postural change. For this reason, select parameters change after HUT in order to recreate dynamics observed in data. We select which parameters to change based on the phenotype that we are representing. We decrease the upper body arterial compliance for all simulations to account for the redistribution of volume upon HUT. The values of the changed parameters before and after HUT can be seen in Table \ref{table:phenotype params}

\begin{figure}[ht!]
     \centering
     \begin{subfigure}[b]{0.3\textwidth}
         \centering
         \includegraphics[width=\textwidth]{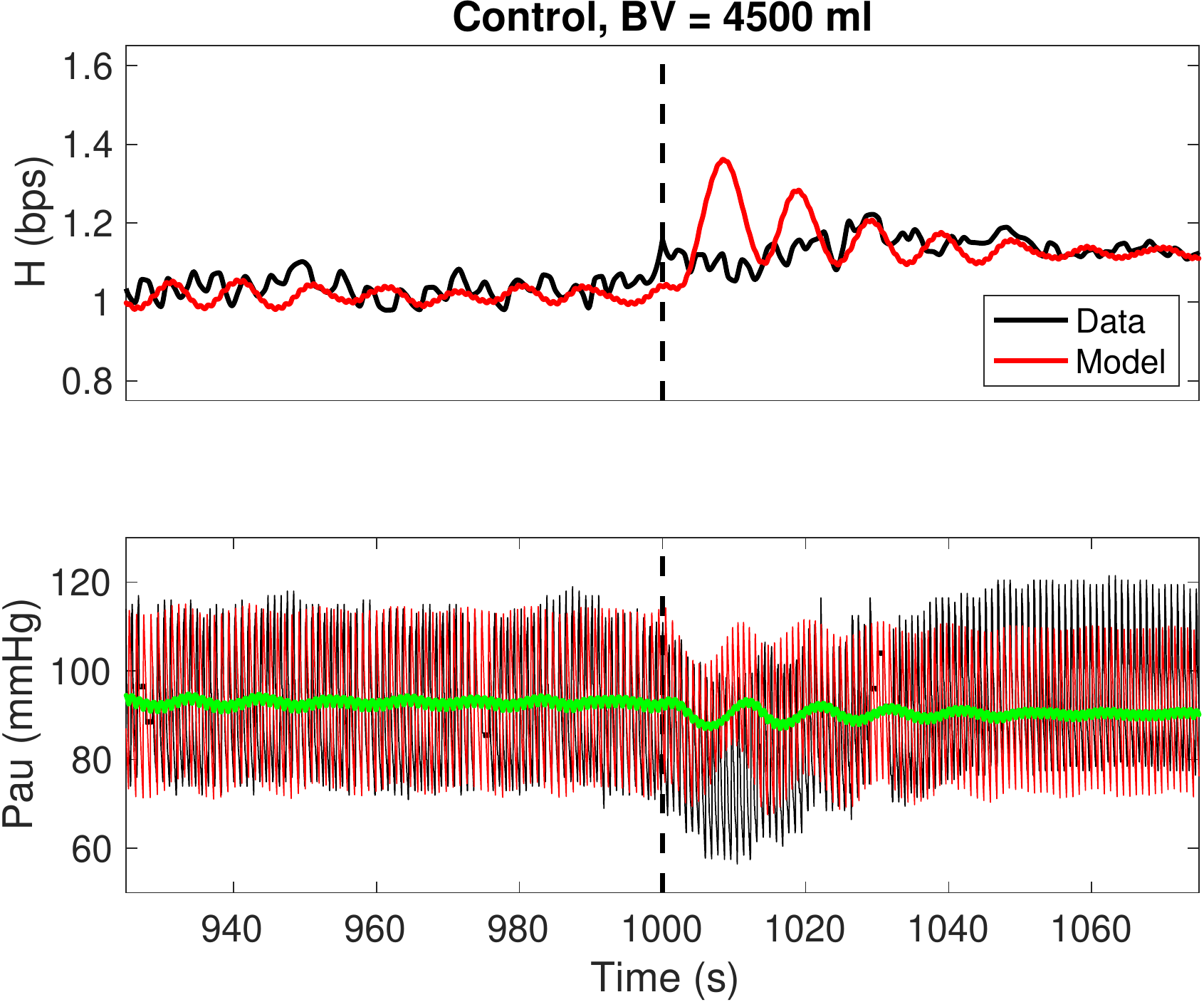}
         \label{fig:rep sub1}
     \end{subfigure}
     \begin{subfigure}[b]{0.3\textwidth}
         \centering
         \includegraphics[width=\textwidth]{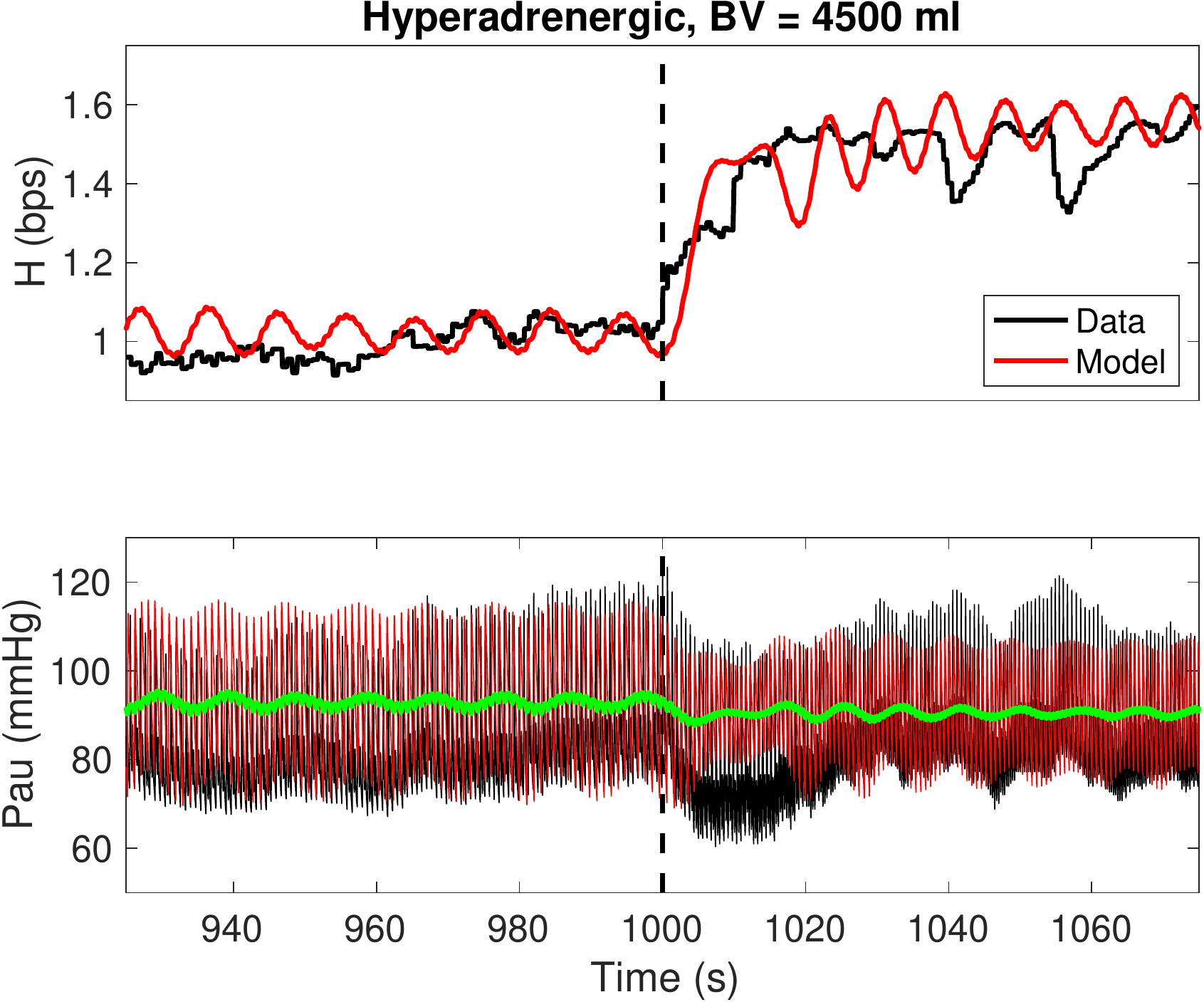}
         \label{fig:rep sub2}
     \end{subfigure}
     \begin{subfigure}[b]{0.3\textwidth}
         \centering
         \includegraphics[width=\textwidth]{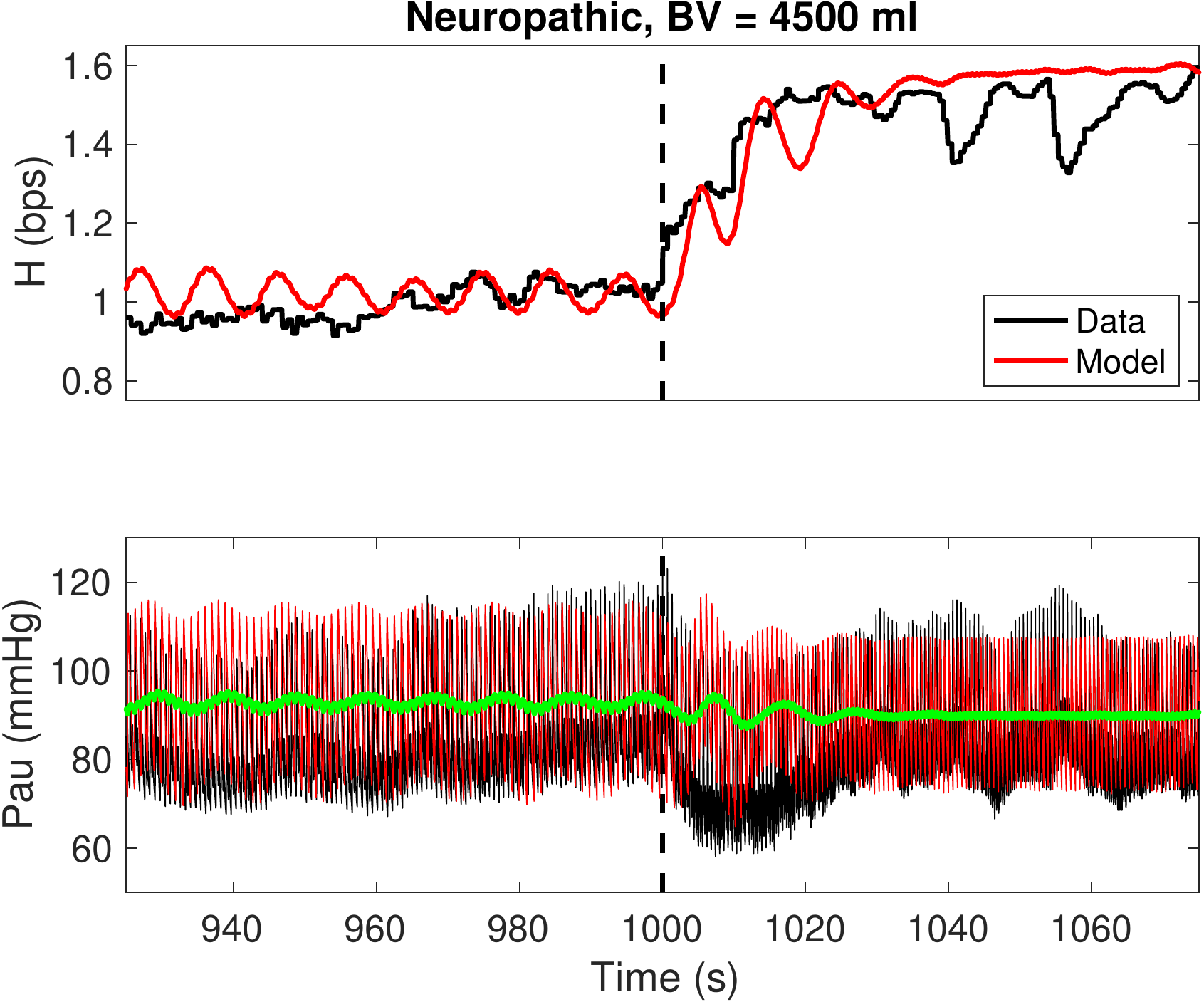}
         \label{fig:rep sub3}
     \end{subfigure}
     \newline
      \begin{subfigure}[b]{0.3\textwidth}
         \centering
         \includegraphics[width=\textwidth]{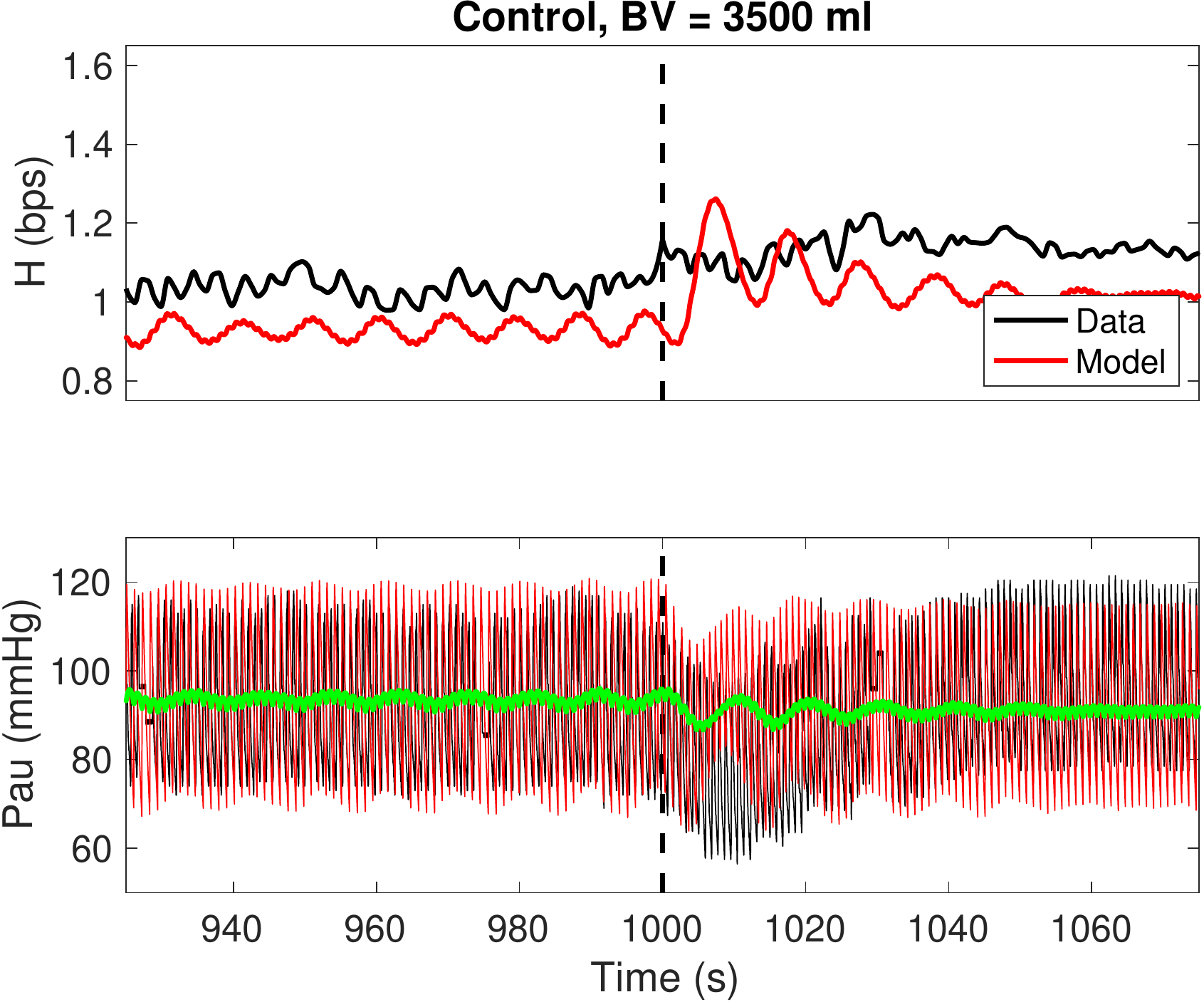}
         \label{fig:rep sub4}
     \end{subfigure}
     \begin{subfigure}[b]{0.3\textwidth}
         \centering
         \includegraphics[width=\textwidth]{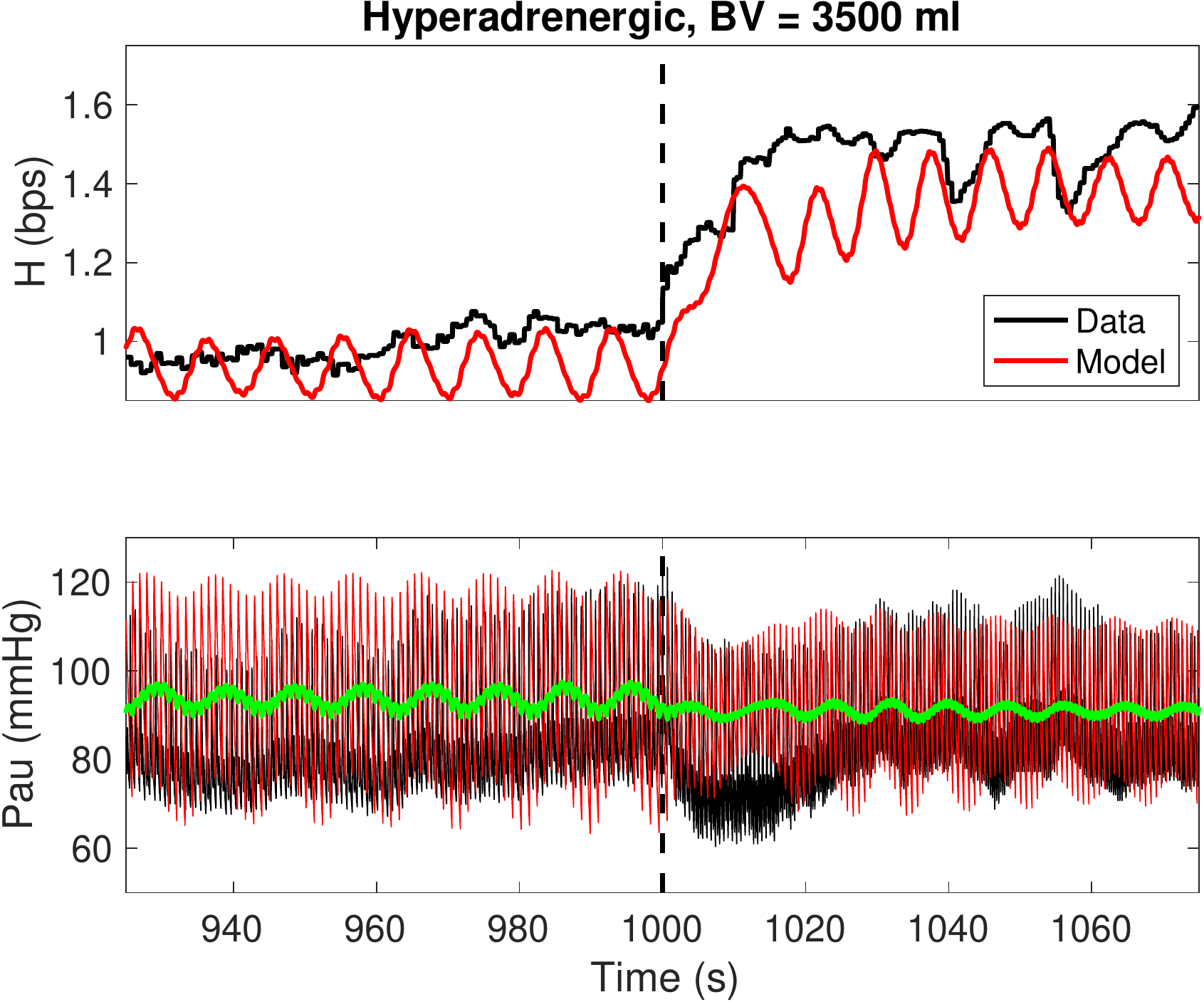}
         \label{fig:rep sub5}
     \end{subfigure}
     \begin{subfigure}[b]{0.3\textwidth}
         \centering
         \includegraphics[width=\textwidth]{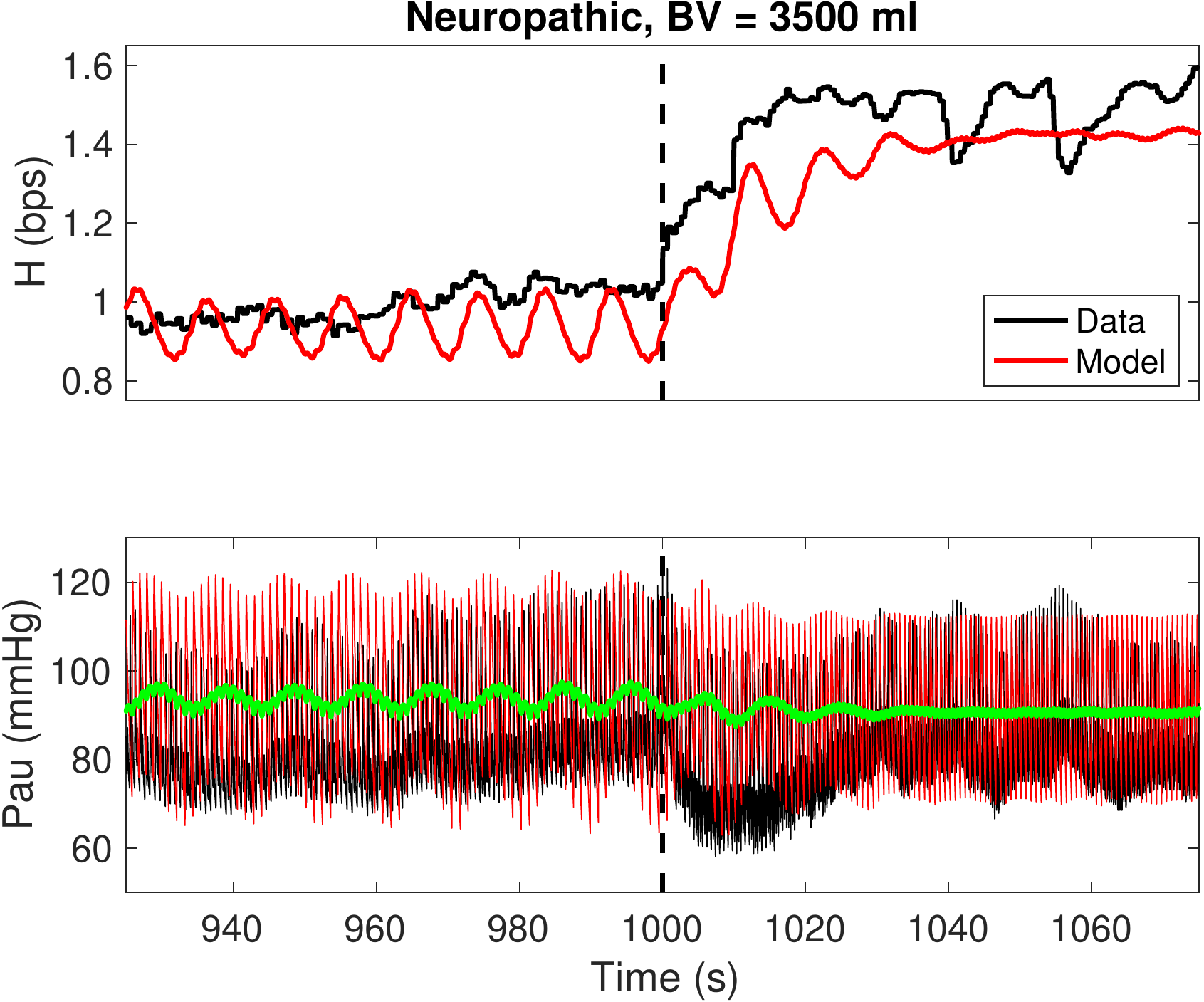}
         \label{fig:rep sub6}
     \end{subfigure}
     \caption{Characteristic data and model predictions of heart rate ($H$), upper arterial blood pressure ($P_{au}$), and mean pressure. Simulations with 4500 ml of blood are in the top row with simulations with 3500 ml of blood in the bottom row. $C_{au}$ is decreased after HUT for all simulations to represent constriction of vasculature upon HUT. In control $P_{2H}$ is decreased (left), $k_H$ and $P_{2H}$ are increased after HUT to replicate hyperadrenergic POTS (middle) and $k_H$ increased, $R_{lpM}$ and $R_{lpm}$ decreased to replicate neuropathic POTS (right).}
     \label{fig:rep pts}
\end{figure}

\paragraph{Control} subjects show a limited increase in heart rate and similar amplitudes of oscillations before and after HUT. When volume is redistributed during HUT, the baroreflex control operating regime is shifted due to upper arterial pressure decreasing. To avoid an increase in heart rate, we shift the heart rate response curve with the pressure by decreasing $P_{2H}$. Simulation of a control subject can be seen with data in the left column of \ref{fig:rep pts}.

\paragraph{Hyperadrenergic POTS} patients have increased levels of plasma norepinephrine during HUT \cite{mar2020postural}. We model this by increasing $k_i$, $i=H,E,R$ and $P_{2H}$ during HUT. Results, depicted in Figure \ref{fig:rep pts} (top row center), show that increasing these parameters increases the amplitude of 0.1 Hz oscillations (compared to the control subject - top left) and causes tachycardia during HUT, which is consistent with POTS patient data.
%JG: Make a table with parameter values as they change for phenotypes

\paragraph{Neuropathic POTS} patients experience excessive blood pooling below the thorax during HUT due to partial autonomic neuropathy. This condition is simulated by decreasing the range of resistance control in the lower body arteries, i.e., we reduce $R_{alp,M}$ and $R_{alp,m}$ making this control less effective. Results from this simulation depicted in Figure \ref{fig:rep pts} top right show that subjects exhibit tachycardia but that oscillations are dampened after HUT onset. 

\paragraph{Hypovolemia} To understand how hypovolemia impacts our predictions, we reduce central blood volume. We found that hypovolemia alone was not able to reproduce POTS dynamics such as increased oscillations or tachycardia. We instead study the effects of hypovolemia on the other phenotypes. Results shown in Figure \ref{fig:rep pts} bottom row show that heart rate is lower compared to patients with a normal blood volume, but for hyperadrenergic POTS patients, the amplitude of the heart rate and blood pressure oscillations increase significantly, indicating that this patient group may experience more severe response to POTS. To study this phenomenon further, we conducted a two-dimensional analysis examining the amplitude of heart rate and blood pressure oscillations as a function of blood volume (BV) and $k_H$ at rest and during HUT. 

Figures \ref{fig:BVkH}A and \ref{fig:BVkH}B (top row) show that reducing blood volume at rest does not impact dynamics. However, as can be seen in the bottom row of Figures \ref{fig:BVkH}A and \ref{fig:BVkH}B, reducing blood volume during HUT increases oscillation amplitude. This implies that more severe oscillations occur during HUT for patients with less blood volume. Similar to Figure \ref{fig:kRkH}, Hopf bifurcation lines can be seen in the parameter space in Figure \ref{fig:BVkH}A but are removed when noise is added to simulations representing heart rate variability as can be seen in Figure \ref{fig:BVkH}B.

\section{Discussion}

This study developed a closed loop baroreflex cardiovascular model and used simple signal processing to extract the frequency and amplitude of heart rate and blood pressure oscillations. Results show that our model can generate oscillations in the low-frequency $\approx 0.1$ Hz) range observed in control and POTS patients at rest and during head-up tilt (HUT) and that oscillations can be manipulated by modulating parameters associated with the baroreflex. 

Our model can predict tachycardia (an increase in HR of at least 30 bpm, 40 bpm in adolescents) observed in POTS patients by increasing the half-saturation of the heart rate response ($P_{2H}$) or decreasing the maximum and minimum vascular resistance ($R_{lpM}$ and $R_{lpm}$). The former is significantly more effective than the latter. Moreover, by changing physiologically relevant baroreflex parameters after HUT, we can reproduce the hyperadrenergic and neuropathic POTS phenotypes suggested by \cite{fedorowski2019postural, mar2020postural}.  Finally, we found that predictions are highly sensitive to changes in blood volume, suggesting that short skinny patients may experience a more severe reaction than females with a normal blood volume.

\subsection{Low-frequency oscillations}

The mathematical model used here extends previous studies \cite{deboer1987hemodynamic, heldt2000computational, williams2014patient, marquis2018practical, matzuka2015using, marquis2018practical, ursino98, ursino2000acute} predicting cardiovascular dynamics using a closed loop lumped parameter model including the left heart, the upper and lower body systemic arteries, and veins. The latter is included to facilitate the redistribution of volume upon postural change. The baroreflex is modeled using a first-order control equation predicting the controlled quantity as a function of pressure using sigmoidal function enforcing saturation at both high and low values of the controlled parameter. 

By modulating parameters associated with the baroreflex sensitivity (the sigmoidal $k_X,X=R,E,H$, we explain the emergence and amplification of the low-frequency oscillations at rest and during HUT. Our findings agree with those reported in our previous study \cite{Geddes2020Osc}, noting that the low-frequency oscillations (sometimes referred to as Mayer waves) are observed in all subjects and that the oscillation amplitude is increased in POTS patients in particular following HUT.  Our findings also agree with previous experimental studies that report larger low-frequency oscillations in cerebral blood flow \cite{medow2014altered, stewart2015oscillatory}.

While this phenomenon has been discussed in studies using signal processing to examine heart rate and blood pressure time-series, only a few studies by Ottesen et al. \cite{ottesen1997modelling} and Ishbulatov et al. \cite{ishbulatov2020mathematical} used closed loop modeling to replicate this phenomenon. Both these studies explained the emergence of oscillations by introducing a delay in sympathetic response. In contrast, our model predicts the emergence of the $\approx 0.1$ Hz oscillatory response without introducing delay differential equations.

These findings agree with the hypothesis that POTS patients may have an abnormally sensitive baroreflex control. Specifically, we observed that $k_H$ is the most influential parameter for the oscillation amplitude. At $k_H<k_{critical}$ , the system does not oscillate, but as $k_H$ increases, oscillations emerge via a Hopf bifurcation. In addition, we found that the baroreflex time-constants modulate the oscillation frequency. 

To better understand how key physiological parameters modulate oscillation amplitude, we conducted a 2-dimensional parameter analysis. Figure \ref{fig:kRkH} shows that during rest and HUT, increased peripheral resistance and heart rate response sensitivities ($k_R,k_H$) increase oscillation amplitude. We see in Figure \ref{fig:kRkH} that during HUT, increased blood volume decreases oscillation. This agrees with clinical insights from Klinik Mehlsen that patients with smaller blood volume have more pronounced POTS symptoms.

\subsection{Head-up tilt (HUT)}

Head-up tilt test is a useful tool for diagnosing POTS \cite{bryarly2019postural}. As a patient is passively tilted up, blood is redistributed due to gravity while the patient is not actively using their muscles to change position. This passive tilt allows a clear depiction of how a patient's body responds to a redistribution of blood volume. Mathematically, we represent the tilt by accounting for gravitational pooling of blood in the lower body as a function of the tilt angle. 

A few modeling studies \cite{heldt2000computational, ishbulatov2020mathematical, williams2014patient} have examined the response to HUT. Williams et al. \cite{williams2014patient} used an open-loop patient-specific model to predict arterial blood pressure using heart rate as an input while Heldt et al.  \cite{heldt2000computational} used a closed-loop cardiovascular model with set-point representations of the baroreflex simulating HUT by increasing pressures in venous compartments, and Ishbulatov et al. \cite{ishbulatov2020mathematical} by increasing pressure to the lower body baroreceptors. The study by Williams et al. \cite{williams2014patient} did not examine low-frequency oscillations and the study by Heldt et al. \cite{heldt2000computational} the low-frequency oscillations were dampened in less than one minute after the onset of HUT, Ishbulatov et al. \cite{ishbulatov2020mathematical} successfully recreated low-frequency oscillations after HUT but only considered healthy subjects. While these studies were able to predict the HUT response, our model is the only one that can generate closed-loop stable oscillations that agrees with POTS patient data.
%JG: JESPER HAD A QUESTION WHAT R
Specifically, we observed that like rest, low-frequency oscillations exist and persist during HUT. However, to get adequate pulse pressure and oscillation amplitude it is necessary to decrease upper arterial compliance to account for the constriction of vasculature upon HUT. We hypothesize that this impact on can be explained by the pressure and volume redistribution. We allow select parameters to change after HUT to duplicate patient data depending on the POTS phenotype appropriately.

\subsection{POTS phenotypes}

The pathophysiology of POTS is complex and not completely understood. Several recent studies \cite{bryarly2019postural, fedorowski2019postural, mar2020postural} speculate that POTS comprise multiple phenotypes including hyperadrenergic, neuropathic, and hypovolemic POTS. Without clearly denoting how these manifest changes in HR and BP time-series, several hypothesis describing each phenotype have been put forward. Hyperadrenergic POTS is believed to be a result of increased levels of circulating norepinephrine, while patients with neuropathic POTS have partial neuropathy in lower vascular beds. Finally, hypovolumic POTS is simply described as POTS in patients with low blood volume.  Additionally, autoantibodies against $\beta_1, \beta_2, \alpha_1,M_1,M_2$ receptors may be responsible for some cases of POTS \cite{fedorowski2017antiadrenergic, fedorowski2019postural}.

In addition to analysis of oscillations we also simulate the two main phenotypes and study how BP and HR change in patients with normal and low blood volume. To predict hyperadrenergic POTS, we increase $P_{2H}$ and $k_X$, $X=H,E,R$ after HUT representing the increased plasma norepinephrine concentration during HUT. In the neuropathic case, we decrease the maximum and minimum response of the lower peripheral resistance ($R_{lpM}, R_{lpm}$) to represent neuropathy in lower extremities \cite{bryarly2019postural}. We observe that increasing $P_{2H}$ in the hyperadrenergic case and reducing $R_{lpM},R_{lpm}$ in the neuropathic case are essential to the presence of orthostatic tachycardia while increasing $k_X,X=H,E,R$ in the hyperadrenergic case is vital to the amplitude of low-frequency oscillations. Figure \ref{fig:rep pts} shows that oscillations are minimal in the neuropathic phenotype. This motivates future work to examine whether all POTS phenotypes exhibit increased low-frequency oscillations in heart rate and blood pressure or if large oscillations are unique to the hyperadrenergic phenotype. 

We were unable to reproduce the dynamics observed in POTS patient data by decreasing blood volume alone, which corresponds to the hypovolemic phenotype put forth by (mar2020postural). Figure \ref{fig:rep pts} also shows that while low blood volume is not able to reproduce POTS dynamics from a control simulation, it can make POTS dynamics more pronounced in simulations where the dynamics are already present. Figure \ref{fig:BVkH} shows that lower blood volume can result in larger oscillations during HUT. These findings imply that hypovolemia may not be a distinct phenotype but rather exacerbates other phenotypes.

\subsection{Heart rate variability (HRV)}

Several previous studies \cite{goldberger1991normal, li2019more, shaffer2017overview} have addressed the importance of heart rate variability. While there is still discussion on the origin of short-term heart rate variability \cite{shaffer2017overview}, the net effect appears as noise. In this study, we accounted for heart rate variability by adding noise to the predicted heart rate. The addition of heart rate variability stabilizes predictions eliminating frequent Hopf bifurcation lines seen in the top row of Figures \ref{fig:kRkH} and \ref{fig:BVkH}. The benefits of added noise in dynamic systems with stable fixed points have been shown before in \cite{breeden1990noise}.

\subsection{Importance of study}
 
This is the first study that uses a closed-loop model of the baroreflex response to explain the oscillatory dynamics observed in POTS patients. Our results advance previous models have had success representing baroreflex responses but with the input of arterial blood pressure \cite{hammer2005resonance, matzuka2015using, williams2014patient,  williams2019optimal}. Others have used a closed-loop model but have used a more complicated baroreflex model and did not specifically consider POTS \cite{ishbulatov2020mathematical}. 

This work holds clinical significance as the POTS phenotypes can be encoded into the model. Therefore, our study provides support for the current hypothesized mechanisms of POTS. We were able to show that hypovolemia contributes to more severe oscillations, which could be linked to more severe symptoms when combined with the other phenotypes. However, we were unable to recreate POTS dynamics by decreasing blood volume alone. We also observed that encoding the neuropathic phenotype resulted in tachycardia upon HUT but not increased oscillations. We were able to successfully recreate observed dynamics by encoding the hyperadrenergic phenotypes into the model.
%JG NOTE BOTH PARAGRAPHS START WITH THIS - REWORD ONE OF THEM

\subsection{Limitations}

Limitations of this work include the oversimplification of the baroreflex model. The baroreflex is a complex negative feedback loop with numerous components, including the baroreceptors, afferent nerves, the nucleus tractus solitarius (NTS) located in the medulla oblongata, efferent nerves, and the actual cell response in the sinoatrial node as well as muscle cells. Lumping these components into four control equations is a large assumption but is done to show that oscillations can be produced even with a simple model. This model aims to provide a simple mathematical formulation to explore the possible origins of POTS. However, more intricate models that accurately represent the true physiology are needed for further exploration. 

Furthermore, we cannot predict the drop in arterial blood pressure immediately after HUT as observed in the data. We also note that the data shown is exemplary and did not attempt to estimate parameters based on this data. This oversimplification hinders the model in predicting the precise hypothesizes of the origin of POTS, such as the exact type of hyperadrenergic antibodies.   Finally, since we did not include a delay, we were not able to recreate the phase differences seen in \cite{Geddes2020Osc}, which are an essential difference between POTS and control patients. Future work will contain a more in-depth description of the baroreflex to replicate these phenomena.

\section{Conclusions}
We have presented a closed-loop differential equation model of the interactions between the baroreflex and cardiovascular system, emphasizing the emergence and amplitude of oscillations in the ~ 0.1 Hz frequency range. We have concluded that the heart rate and peripheral resistance response, represented by Hill coefficients $k_H$ and $k_R$ respectively, and total blood volume are critical to the amplitude of low-frequency oscillations. Results shared here help explain clinical observations and motivate further modeling and study of POTS to understand better the disease's pathophysiological aspects and possible treatment options.

\section*{Acknowledgment}
We acknowledge and thank Francis Polakiewicz for his help in developing the model and preliminary analysis.

\section*{Funding}
JG has been funded by the National Science Foundation under the award NSF/DMS(RTG) \#1246991 and NSF/DMS \#1638521

% \bibliography{bibPOTS}
\printbibliography

\end{document}